\documentclass{aa}  

\usepackage{longtable}
\usepackage{graphicx}
\usepackage{color}

\usepackage{txfonts}
\begin{document} 

   \title{Observational study of intermittent solar jets: p-mode modulation}
   
    \author{Qiuzhuo Cai \inst{1} 
             \and 
             Guiping Ruan \inst{1} 
             \and 
             Chenxi Zheng \inst{1} 
             \and
             Brigitte Schmieder \inst{2,3} 
             \and  
             Jinhan Guo \inst{2,4}
             \and  
             Yao Chen \inst{1}
            \and 
            Jiangtao Su \inst{5,6} 
            \and  
            Yang Liu  \inst{1} 
            \and
            Jihong Liu  \inst{7}
            \and
            Wenda Cao   \inst{8,9}  
            }

\institute{Shandong Provincial Key Laboratory of Optical Astronomy and Solar-Terrestrial Environment, and Institute of Space Sciences, Shandong University, Weihai 264209, China\\
            \email{rgp@sdu.edu.cn}
            \and 
             Centre for Mathematical Plasma-Astrophysics, Department of Mathematics, KU Leuven, Celestijnenlaan 200B, 3001 Leuven, Belgium 
             \and 
             LESIA, Observatoire de Paris, Universit\'e PSL, CNRS, Sorbonne Universit\'e, Universit\'e de Paris, 5 place Jules Janssen, 92190 Meudon, France
             \and 
              School of Astronomy and Space Science and Key Laboratory for Modern Astronomy and Astrophysics, Nanjing University, Nanjing 210023, China
             \and 
             National Astronomical Observatories, Chinese Academy of Sciences, Beijing, 100101,China
              \and 
              School of Astronomy and Space Sciences, University of Chinese Academy of Sciences,Beijing 100049, China
             \and 
             Shi Jiazhuang University, Shi Jiazhuang 050035, China
             \and    
             Center for Solar-Terrestrial Research, New Jersey Institute of Technology, 323 Martin Luther King Blvd., Newark, NJ 07102, USA
            \and    
            Big Bear Solar Observatory, 40386 North Shore Lane, Big Bear City, CA 92316, USA
   }

 
  \abstract
   {}
   {Recurring jets are observed in the solar atmosphere. They can erupt intermittently over a long period of time. By the observation of intermittent jets, we wish to understand what causes the characteristics of the periodic eruptions.}
  {We report intermittent jets observed by the Goode Solar Telescope (GST) with the TiO Broadband Filter Imager (BFI), the Visible Imaging Spectrometer (VIS) in H$_\alpha$, and the Near-InfraRed Imaging Spectropolarimeter (NIRIS). The analysis was aided and complemented by  1400~\AA{} and 2796~\AA{} data from the Interface Region Imaging Spectrograph (IRIS). These observational instruments allowed us to 
  analyze the temporal characteristics of the jet events. By constructing the H$_\alpha$ Dopplergrams, we found that the plasma first moves upward, but during the second phase of the jet, the plasma flows back. Working with time slice diagrams, we investigated the 
  characteristics of the jet dynamics.}
   {The jet continued for up to 4 hours. The time-distance diagram shows that the peak of the jet has clear periodic-eruption characteristics (5 minutes) during 18:00 UT-18:50 UT. We also found a periodic brightening phenomenon (5 minutes) during the jet bursts in the observed bands in the transition region (1400~\AA{} and 2796~\AA{}), which may be a response to intermittent jets in the upper solar atmosphere. The time lag is 3 minutes. Evolutionary images in the TiO band revealed a horizontal movement of the granulation at the location of the jet. By comparison to the quiet region of the Sun, we found that the footpoint of the jet is enhanced at the center of the H$_\alpha$ spectral line profile, without significant changes in the line wings. This suggests prolonged heating at the footpoint of the jet. In the mixed-polarity magnetic field region of the jet, we observed the emergence of magnetic flux, its cancellation, and shear, indicating possible intermittent magnetic reconnection. This is confirmed by the nonlinear force-free field model, which was reconstructed using the magneto-friction method. }
   {The multiwavelength analysis indicates that the events we studied were triggered by magnetic reconnection that was caused by mixed-polarity magnetic fields. We suggest that the horizontal motion of the granulation in the photosphere drives the magnetic reconnection, which is modulated by p-mode oscillations.}


   \maketitle
%

\section{Introduction}
Solar jets are plasma ejection phenomena that are observed throughout the solar atmosphere and have been extensively studied in terms of their morphology, dynamic characteristics, and driving mechanisms since their first detection in the X-ray emission of coronal 
jets by the Soft X-ray Telescope on board the Yohkoh satellite in the early 1990s \citep{Schmieder1995,Shen2019,Raouafi2016,Shen2021,Schmieder2022F}. 
Jets are observed in multiple wavelengths in the solar atmosphere. They appear as bright structures in the corona and as plasma flows along 
magnetic field lines in the chromosphere \citep{Tian2018,De2021,Schmieder2022}. They have been referred to as H$_\alpha$ surges, plasma ejections, and chromospheric jets in previous studies \citep{Roy1973,Asai2001,Louis2014}. Recently, some researchers have also called them light walls \citep{Yang2015} or peacock jets \citep{Robustini2016}. \citet{Zhao2022} analyzed recurrent jets that repeatedly propagated from one end to the other in the chromosphere. Many jets have been observed in sunspots, while others occur in light bridges, such as the fan-shaped jets near sunspot light bridges studied by \citet{Liu2022}. The first observation of fan-shaped jets on sunspot light bridges was reported by \citet{Asai2001}, who found that these jets had speeds of about 50~km/s and a maximum length of 2 mega meters, suggesting that the jets originate from emerging magnetic flux with no compelling observational evidence.

Jets can occur above neutral lines of magnetic fields \citep{Hou2016} and are thought to be triggered by magnetic reconnection, either in combination with magnetic acoustic waves \citep{Zhang2017}, magnetic reconnection \citep{Hou2017,Bai2019,Yang2019}, or a combination of both \citep{Tian2018,Huang2020}. Magnetic reconnection is widely considered as the triggering mechanism for jets, and researchers have been searching for evidence of magnetic reconnection in the solar atmosphere. Some high-resolution observations have shown inverted-Y-shaped jets that frequently occur in coronal holes and active regions around sunspots \citep{Cirtain2007,Singh2012,Yang2011,Tian2012,Zhang2014,Shen2012}. Inverted-Y-shaped jets are considered to be the result of reconnection between small-scale magnetic bipolar and unipolar background fields. The observations provide strong evidence for magnetic reconnection \citep{Moreno2013,Chen2015,Tian2018}. The former authors studied reconnection-driven jets that repeatedly occur on the light bridges of sunspots. They examined jets that frequently occurred in the wings of the H$_\alpha$ line and found that many jets exhibited an inverted-Y-shaped structure, demonstrating a typical reconnection process in a unipolar magnetic field environment where the overlying magnetic field of the penumbra reconnected with newly emerged magnetic flux. A wealth of evidence was also reported from numerical simulations that supports the connection between jets and magnetic reconnection. For example, \citet{Yokoyama1995,Yokoyama1996} performed numerical simulations based on the magnetic reconnection model to reproduce coronal X-ray jets, which successfully demonstrated the connection between jets and magnetic reconnection. They generated anemone jets and bidirectional jets in their simulations based on two different initial magnetic field configurations. The anemone jets were produced by reconnection between newly emerged and coronal sheared fields, mostly along the spine \citep{Joshi2020,Zhu2023}. The bidirectional jets were produced by reconnection between newly emerged and overlying fields \citep{Ruan2019}. Both types of jets confirmed the occurrence of magnetic reconnection.

It is generally thought that jets are associated with the emergence and cancellation of magnetic flux. Many observational results support the model based on which magnetic reconnection triggers jet events. The interaction between emerging magnetic flux fields and the mobile magnetic structure can trigger jet events \citep{Brooks2007}. \citet{Kurokawa1993} found that jets were frequently observed and recurred for several hours, leading to the conclusion that magnetic reconnection between the newly emerged flux and preexisting magnetic fields is the basic mechanism for generating jets. \citet{Shimojo1998} studied the magnetic field characteristics of X-ray jets and found that jets occur in unipolar, bipolar, and mixed-polarity regions, highlighting the importance of the magnetic field environment in the occurrence of jets. \citet{Chae1999} analyzed ultraviolet jets in the transition region and found that they repeatedly occur in regions where preexisting magnetic flux of opposite polarity cancels out with newly emerged magnetic flux. \citet{Liu2004} studied a jet event in an emerging flux region and found a close correlation between the jet and the newly emerged bipolar structure, suggesting that an enhanced magnetic cancellation process triggered the jet. \citet{Yoshimura2003} reported a close correlation between a jet at the edge of emerging flux regions, magnetic cancellation, and ultraviolet brightening, indicating a strong spatiotemporal relation between jets and the brightening observed in the photosphere, especially during the early stages of flux emergence, which is consistent with the model of magnetic reconnection. 

In addition to the scenario that magnetic reconnection triggers jets, Magnetohydrodynamic (MHD) waves in the photosphere may also play a role. Magneto-acoustic waves caused by p-mode leakage or Alfvén waves can lead to the formation of shocks, which then propel the plasma into magnetic flux tubes by increasing the magnetic pressure of the giant spicules \citep{Shibata1982}. \citet{Shibata1982} used a one-dimensional (1D) MHD model to explain why the spicules are longer in coronal holes, with the key process being the increased intensity of the chromospheric shock waves. Based on this, \citet{Iijima2015} studied the influence of the coronal temperature on chromospheric jets and found through two-dimensional (2D) MHD simulations that jets are ejected farther outward when the coronal temperature is lower (similar to coronal holes). Subsequently, \citet{Iijima2017} used a three-dimensional (3D) MHD model to study jets that were generated by twisted magnetic field lines and observed the excitation of various MHD waves and the generation of chromospheric jets in their simulations. The strong twisting of magnetic field lines in the chromosphere helps to drive the jets through the action of the Lorentz force, which means that jets are a natural outcome of oscillatory motion.

Repeated jets often occur in mixed-polarity regions \citep{Chen2015,Jiang2000,Guo2013,Joshi2017}, where persistent flux emergence, cancellation, and convergence can lead to the repeated occurrence of jets. Repeating jets often occur in nearly the same location \citep{Schmieder1995,Chifor2008,Zhang2012,Wang2012,Wang2006}. \citet{Cirtain2007} detected an average of ten jet events per hour in a 100-hour observation and found that jets often occurred at the same X-ray bright point or very close to the location of the previous jet onset. \citet{Jiang2007} observed three jet events that occurred intermittently within approximately 70 minutes. Guo et al.\ (2013) reported three recurring extreme-ultraviolet (EUV) jets within an hour and attributed them to repeated accumulated currents. Mulay et al.\ (2017) studied periodic jets using Si IV 1400~\AA{} data obtained from the Interface Region Imaging Spectrograph (IRIS) slit-jaw imager (SJI) and observed bright and compact plasmas, suggesting a helical motion along the apex of the jet. \citet{Yang2015} discovered many bright structures rooted in the light bridges of active region sunspots and named them "light walls." The tops of these bright walls exhibit sustained upward and downward motion that oscillates in height with a period of approximately 4 minutes. They interpreted these oscillations as leakage of p-mode waves from beneath the photosphere.

P-mode oscillations in the photosphere may contribute to periodic solar activities, as shown by \citet{Chandra2015}, who reported recurring jets with an oscillation period of approximately 3 minutes. They suggested that the increase and decrease of the sunspot oscillation power before and after the jet can indicate the occurrence of magnetic reconnection dominated by a wave, and then modulated the 3-minute period of the jet, which might correspond to the leakage of 3-minute slow magnetoacoustic waves.  
Recently, 3D numerical MHD simulations of a model solar atmosphere with a uniform, vertical, and cylindrically symmetric magnetic field, mimicking the behavior of p-mode oscillations were performed in a pore \citep{Griffiths2023}. The authors concluded that the magnetic regions of the solar atmosphere are favorable for the propagation of a small leakage of energy by slow magnetosonic modes. It was found that the oscillations are enhanced by a vertical magnetic field. The results also exhibit a variation in the frequency of the oscillations at different heights in the low to medium solar atmosphere and for different values of the magnetic field.

\citet{Zeng2013} found a recurring jet with a 5-minute period in their previous study. \citet{Hong2022} studied quasi-periodic microjets driven by granulation convection and proposed that the persistent cancellation of opposite-polarity magnetic flux triggered by p-mode oscillations from the solar interior modulates the possibly intermittent magnetic reconnection and thus controls the 5-minute periodicity of the jets. The modulation of magnetic reconnection by p-mode oscillations is more likely to occur in small-scale, low jet events beneath the chromosphere \citep{Chen2006,Hansteen2006,De2007}, and further observational evidence is expected to support this.

In this paper, we present high-resolution observations of  intermittent jets obtained by the Goode Solar Telescope (GST) operating at the Big Bear Solar Observatory (BBSO), as well as observations by the Solar Dynamic Observatory (SDO) coupled with the Helioseismic and Magnetic Imager (HMI) and the Interface Region Imaging Spectrograph (IRIS). We analyze the dynamic characteristics of  the intermittent jets, the line profile, and the Doppler velocity at the footpoints of jets. Additionally, we analyze the transition region brightening phenomenon and the magnetic field environment of the jets  studied by a nonlinear force-free-field (NLFFF) analysis in Section~2. In Section~3 we summarize our results and discuss the influence of granular motions on the triggering mechanism of intermittent jets.

\begin{figure*}[!htbp]
\begin{minipage}{\textwidth}
\centering
\includegraphics[width=170mm,angle=0,clip]{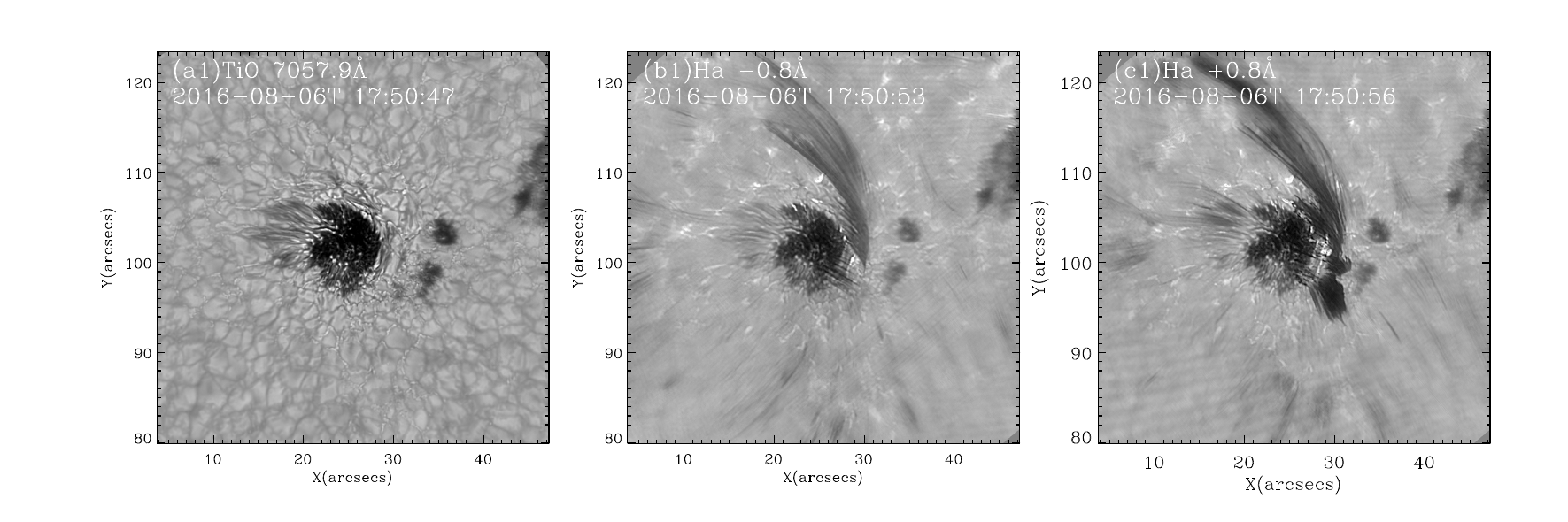}
\includegraphics[width=170mm,angle=0,clip]{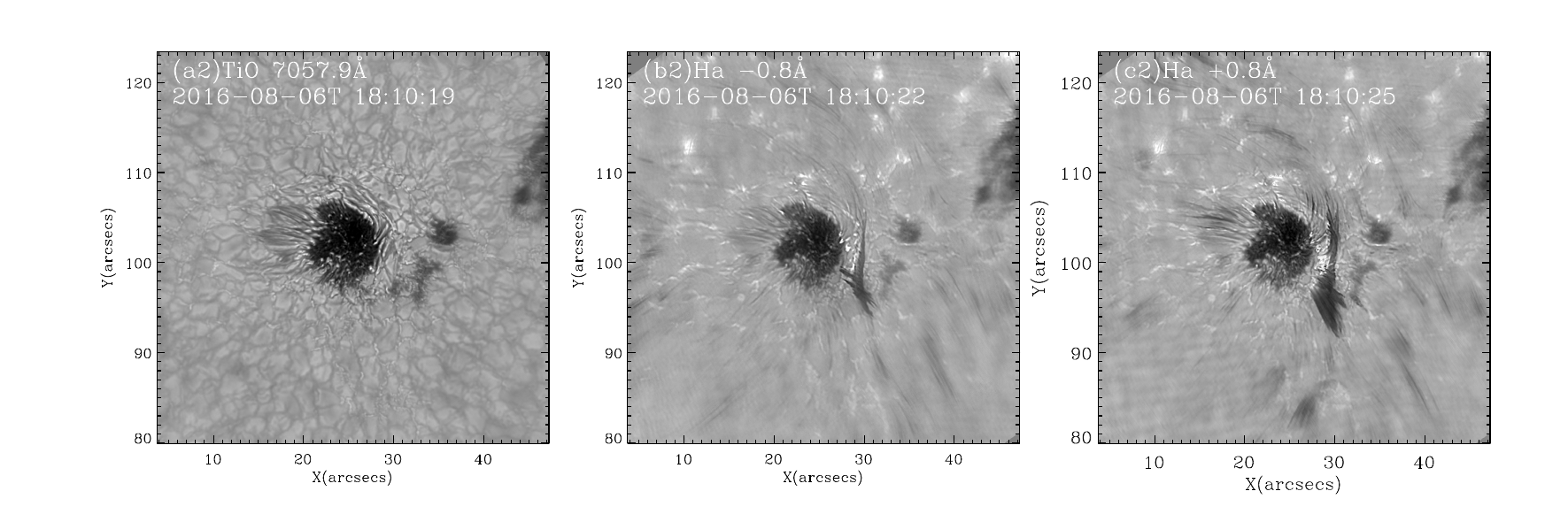}
\includegraphics[width=170mm,angle=0,clip]{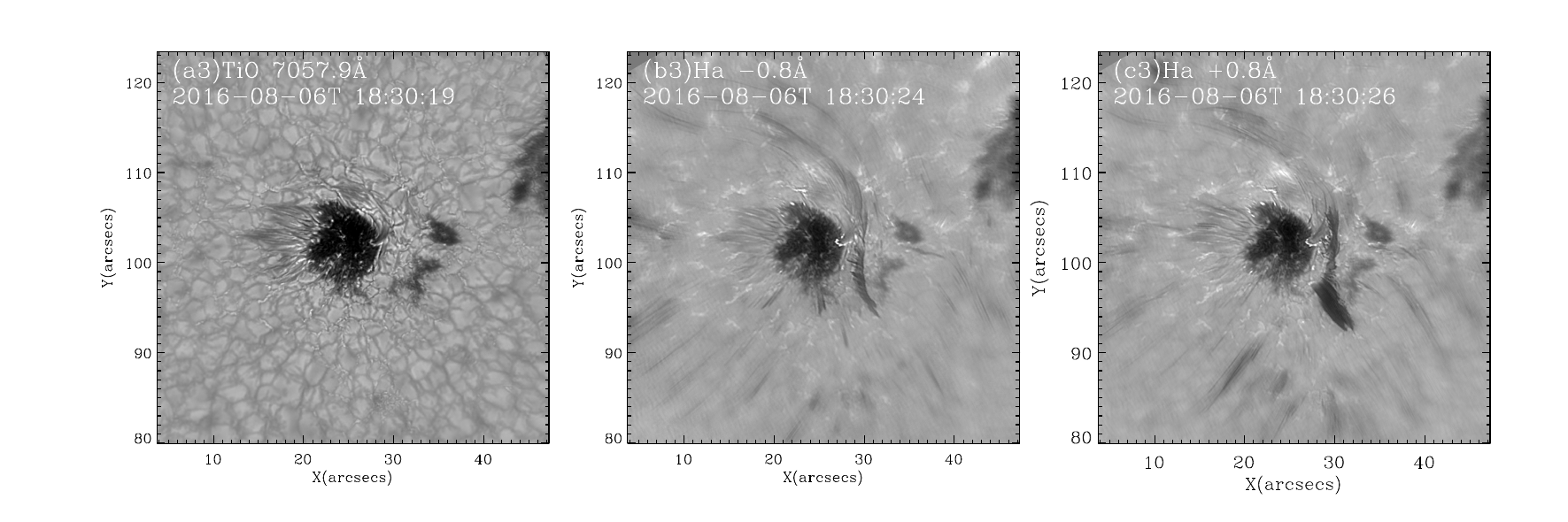}
\includegraphics[width=170mm,angle=0,clip]{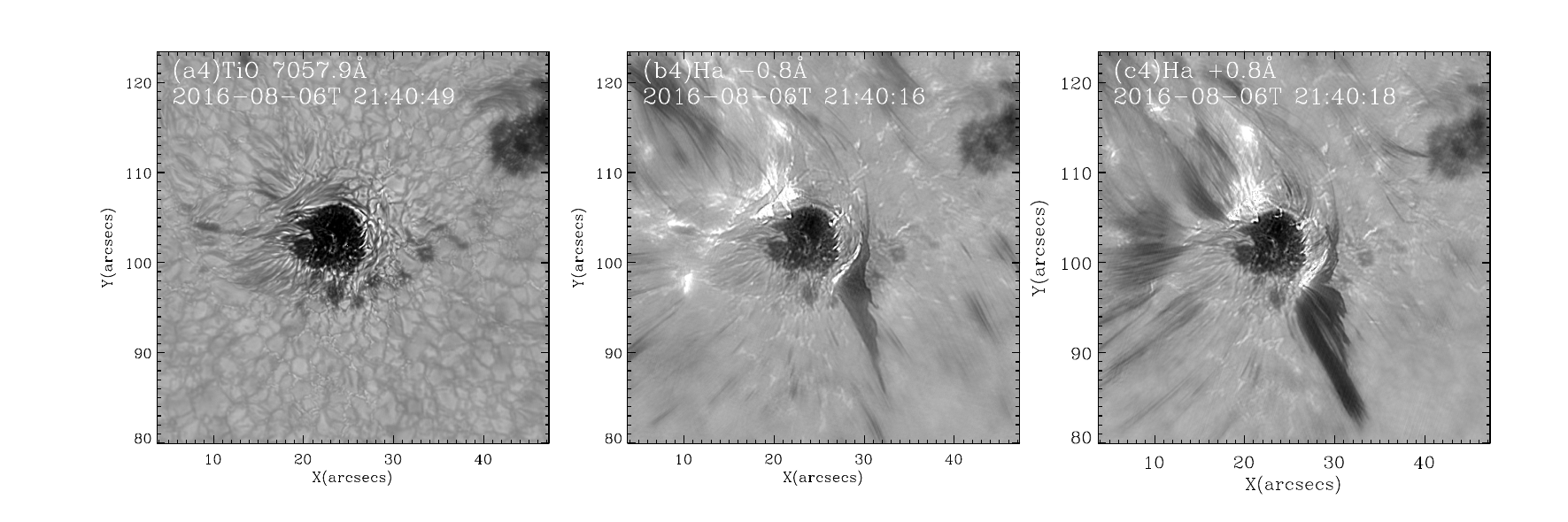}
\end{minipage}

\caption{ Temporal evolution of intermittent jets observed in active region NOAA  12571 with the GST at BBSO on August 6, 2016, between  17:50 UT and 21:40 UT. The different panels show the TiO images (panels a1, a2, a3, and a4), the H$_{\alpha}$ blue-wing images at -0.8 \AA{} (panels b1, b2, b3, and b4), and the H$_{\alpha}$ red-wing images at +0.8 \AA{}   (panels c1,c2, c3, and c4).  All H$\alpha$ images are  obtained with the GST/VIS.}
\label{figure_1}
\end{figure*}


\section{{Observations}}
\label{observation}
\subsection{Instruments}
On August 6, 2016, intermittent jets were observed 
between the two main sunspots of negative polarity that formed  the leading part of  NOAA 
AR 12571 located at N13W05 with the Big Bear Solar Observatory (BBSO) coupled with the 1.6-meter Goode Solar Telescope (GST) \citep{Goode2012} as well as with the Solar Dynamic Observatory (SDO) \citep{pesnell2012} coupled with the Helioseismic and Magnetic Imager (HMI) \citep{Scherrer2012,Schou2012} and the Interface Region Imaging Spectrograph (IRIS) \citep{Depontieu2014}. The pointer of the GST was centered on the eastern sunspot in the  leading polarity of the active region.

The GST data contain simultaneous observations of the photosphere, using the titanium oxide (TiO) line taken with the Broadband Filter Imager, and the chromosphere, using the H$_{\alpha}$ 6563~\AA{} line obtained with the Visible Imaging Spectrometer (VIS) \citep{cao2010}. The passband of the TiO filter is 10~\AA{}, centered at 705.7~nm, and its temporal resolution is about 15 s with a pixel scale of 0.$^{\prime\prime}034$. A combination of a 5~\AA{} interference filter and a Fabry-P{\'e}rot \'etalon  is used at the VIS to obtain a bandpass of 0.07~\AA{} in the H$_{\alpha}$ line. The VIS field of view (FOV) is about 70$^{\prime\prime}$ with a pixel scale of $0.^{\prime\prime}029$. To obtain more spectral information, we scanned the H$_{\alpha}$ line at five positions with a step of 0.4\AA{} following this sequence: $\pm$ 0.8, $\pm$ 0.4, 0.0 \AA{}. We obtained a full Stokes spectroscopic polarimetry using the Fe I 1565~nm doublet over a 85$^{\prime\prime}$ round FOV with the aid of a dual Fabry-P{\'e}rot etalon by the NIRIS Spectropolarimeter. Stokes I, Q, U, and V profiles were obtained every 72 s with a pixel scale of 0.$^{\prime\prime}$081. All TiO and H$_{\alpha}$ data were speckle reconstructed using the Kiepenheuer-Institute Speckle Interferometry Package \citep{woger2008}.

We first analyzes the vector magnetic field and continuum intensity data  given by HMI.
Generally, HMI provides four main types of data: dopplergrams (maps of solar surface velocity), continuum filtergrams (broad-wavelength photographs of the solar photosphere), and both line-of-sight and vector magnetograms (maps of the photospheric magnetic field). The processed HMI continuum intensities and magnetograms data are obtained with a 45~\rm{s} cadence and a size of $ 0.^{\prime\prime}6 $ pixels provided by the HMI team. For comparison with NIRIS, we analyzed the HMI magnetograms in the 24 hours before the event. Continuum-intensity maps of HMI help us to co-align the TiO and H$_{\alpha}$ images and the magnetograms taken by GST. The GST images taken at each wavelength position were internally aligned using the cross-correlation technique provided by the BBSO programmers. 

IRIS provides ultraviolet images focused on the three main channels of the IRIS telescope. Their corresponding wavelengths are far-ultraviolet short (FUVS; 1331.56\AA{}–1358.40\AA{}), far-ultraviolet long (FUVL; 1390.00\AA{}–1406.79\AA{}), and near-ultraviolet (NUV; 2782.56\AA{}–2833.89\AA{}). Wideband filters in CCD imaging provide the imaging data for the Slit-Jaw Imager (SJI). IRIS observed in the sit-and-stare spectral mode with one slit located just at the northern edge of the eastern leading sunspot, just in the middle of the FOV of GST.  In this study, we used the image from the SJI channels at 2832\AA{}, 2796\AA{}, and 1400 \AA{} and the Si IV spectra (see section 3). The coalignment between HMI continuum and GST/IRIS images was achieved by comparing commonly observed features of sunspots in Fe I 6173\AA{} images and TiO and H$_{\alpha}$ $\pm$ 0.8\AA{} images taken frame by frame.

 \subsection{Intermittent jets}
 \label{Intermittent jets}
We are interested in intermittent jets in the leading spot of AR 12571 observed in the H$_{\alpha}$ lines with the GST. 
According to the sunspot whorls, the helicity of the active region is positive. 
The event occured between 17:50:53 UT and 21:40:16UT and lasted for 4 hours. Figure~\ref{figure_1} shows the temporal evolution of the event for three wavelengths: TiO, H$_{\alpha}$ -0.8 \AA{}, and H$_{\alpha}$+ 0.8 \AA{} at four different times, indicating the occurrence of jets throughout observations of nearly 4 hours. From the H$_{\alpha}$ blue-wing (-0.8 \AA{}) images, it is evident that the jets predominantly erupted from the right side of the observed spot. The evolution of the images in the TiO show compression motion in the granulation at the jet location, which may trigger magnetic reconnection.

\begin{figure*}[!htbp]
\begin{minipage}{\textwidth}
\centering
\includegraphics[width=170mm,angle=0,clip]{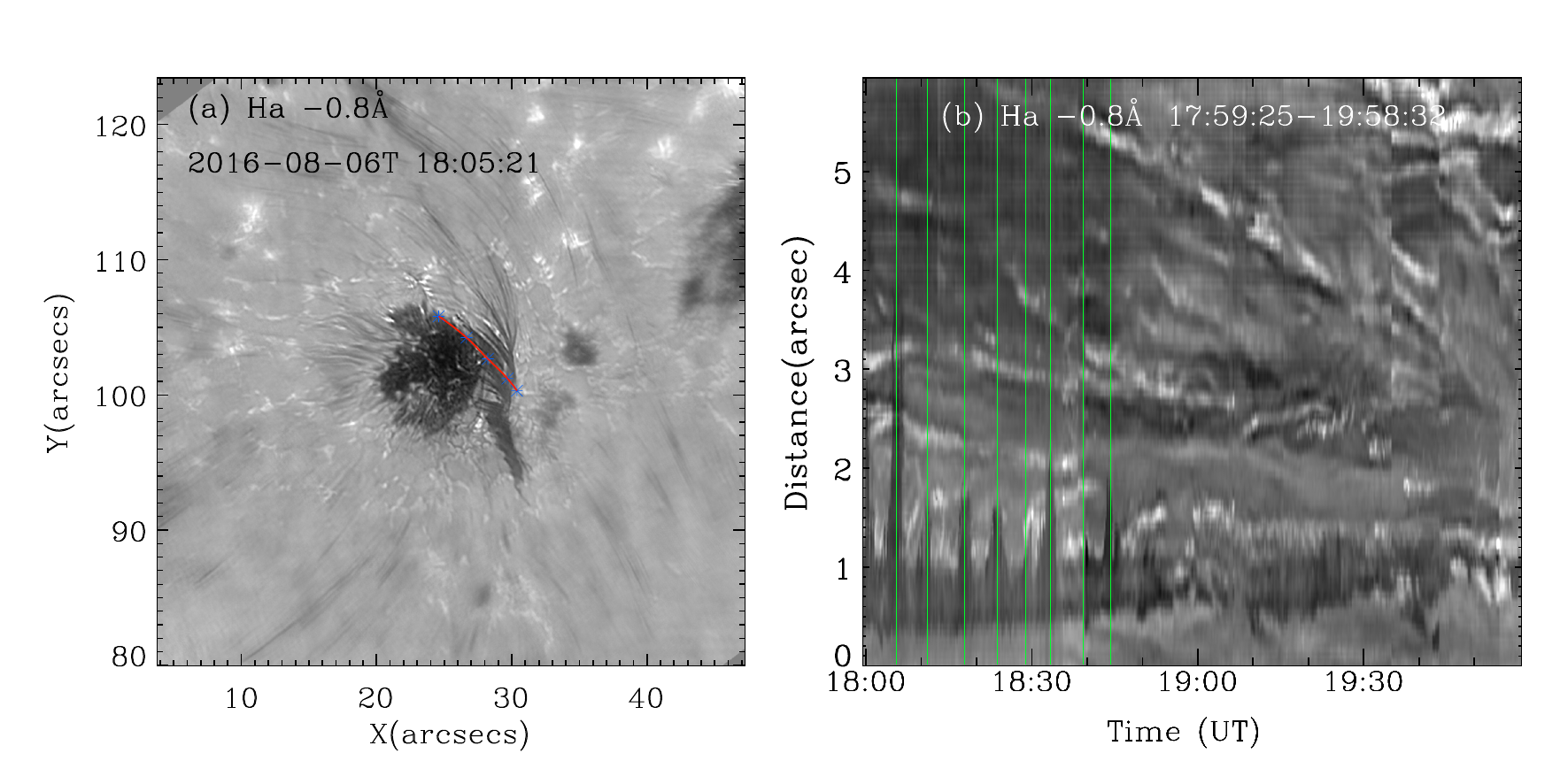}
\end{minipage}
\caption{Oscillation behavior shown in the time-slice diagrams in S1 (red curve) marked in  the H$_{\alpha}$ blue-wing image at -0.8 \AA{} (left panel). The right panels show the plasma trajectories moving along S1. S1 represents the motion path of the intermittent jet.}
\label{figure_2}
\end{figure*}

Using the observation data from the H$_{\alpha}$ blue wing (-0.8 \AA{}), we selected the red curve shown in Figure~\ref{figure_2} (a) as the slice position (referred to as S1) and obtained a time-distance diagram of the jet eruption from 17:59:25 UT to 19:58:32 UT (Figure~\ref{figure_2}(b)). It can be observed that the jet continuously erupted at varying heights. Interestingly, between 18:00:00 UT and 18:50:00 UT, the time-distance diagram of the jets shows a certain periodicity. In Figure~\ref{figure_2}(b), we marked the peak time of jet eruption height with green vertical lines, and it was found that the jet erupted approximately every 5 minutes during this period. This periodicity is likely related to the magnetic field environment in which the jet is situated.

 \subsection{Transition region response}
 \label{Transition region response}
\begin{figure*}[!htbp]
\begin{minipage}{\textwidth}
\centering
\includegraphics[width=170mm,angle=0,clip]{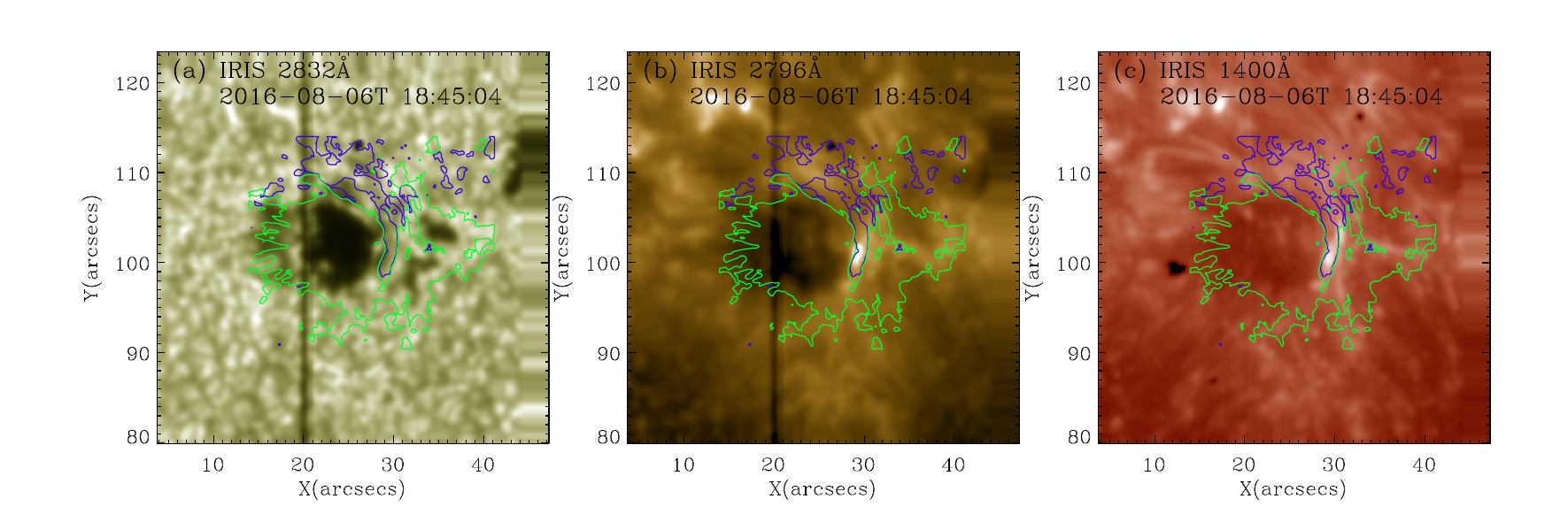}
\end{minipage}
\caption{Images of SJI 2832 \AA{}, 2796 \AA{}, and 1400 \AA{} superimposed with BBSO/NIRIS longitudinal field data. The blue and green lines represent the longitudinal magnetic field boundary contours of +30 and -200 G, respectively.}
\label{figure_3}
\end{figure*}

To investigate the phenomena of intermittent jets in the solar atmosphere above the chromosphere, we obtained slit-jaw imaging (SJI) data from the Interface Region Imaging Spectrograph (IRIS) for three wavelengths: 2832~\AA{}, 2796~\AA{,}, and 1400~\AA{}. The IRIS observation data cover approximately two hours from 17:59:26~UT to 19:58:41~UT. To confirm whether these brightenings correspond to the intermittent jets, we overlaid the longitudinal magnetic field data from GST on the SJI images of the three wavelengths in Figure~\ref{figure_3}. We also identified UV brigthenings in 1400~\AA{} and 2796~\AA{} at the location of the jets. The brightenings precisely correspond to the boundaries of positive and negative magnetic fields, indicating the response of the intermittent jets to brightenings in the chromosphere and transition region. 

\begin{figure*}[!htbp]
\begin{minipage}{\textwidth}
\centering
\includegraphics[width=170mm,angle=0,clip]{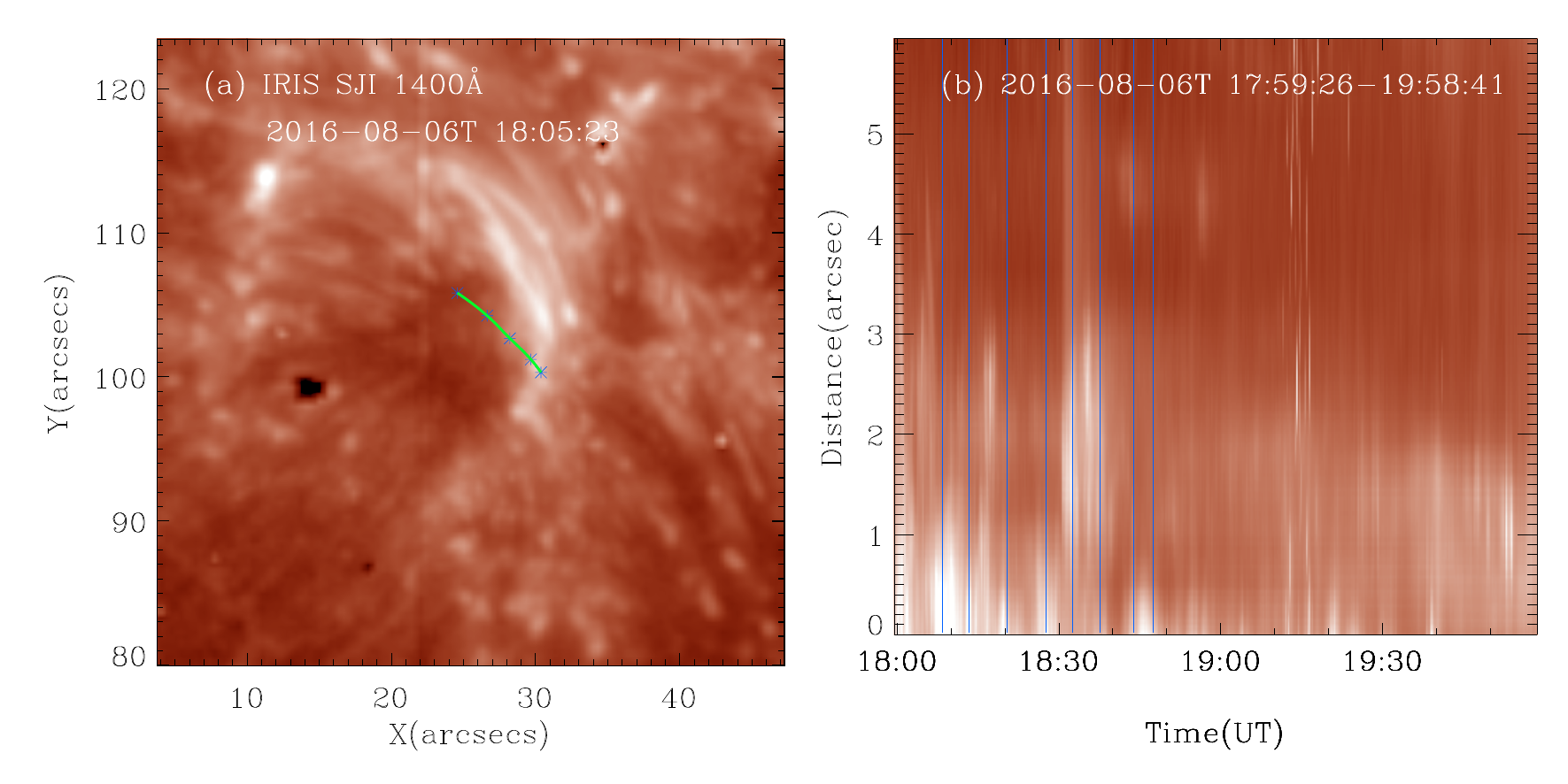}
\end{minipage}
\caption{ Oscillation behavior  shown in the time-slice diagram of the evolution of the intensity at the jet base. Panel (a): IRIS SJI 1400 \AA\ image. The green curve indicates the location  of the slit for the time-slice diagram. Panel (b): Time-slice diagram along the slit. The vertical blue lines were used to mark the moments of clear brightening. }
\label{figure_4}
\end{figure*}

\begin{figure*}[!htbp]
\begin{minipage}{\textwidth}
\centering
\includegraphics[width=170mm,angle=0,clip]{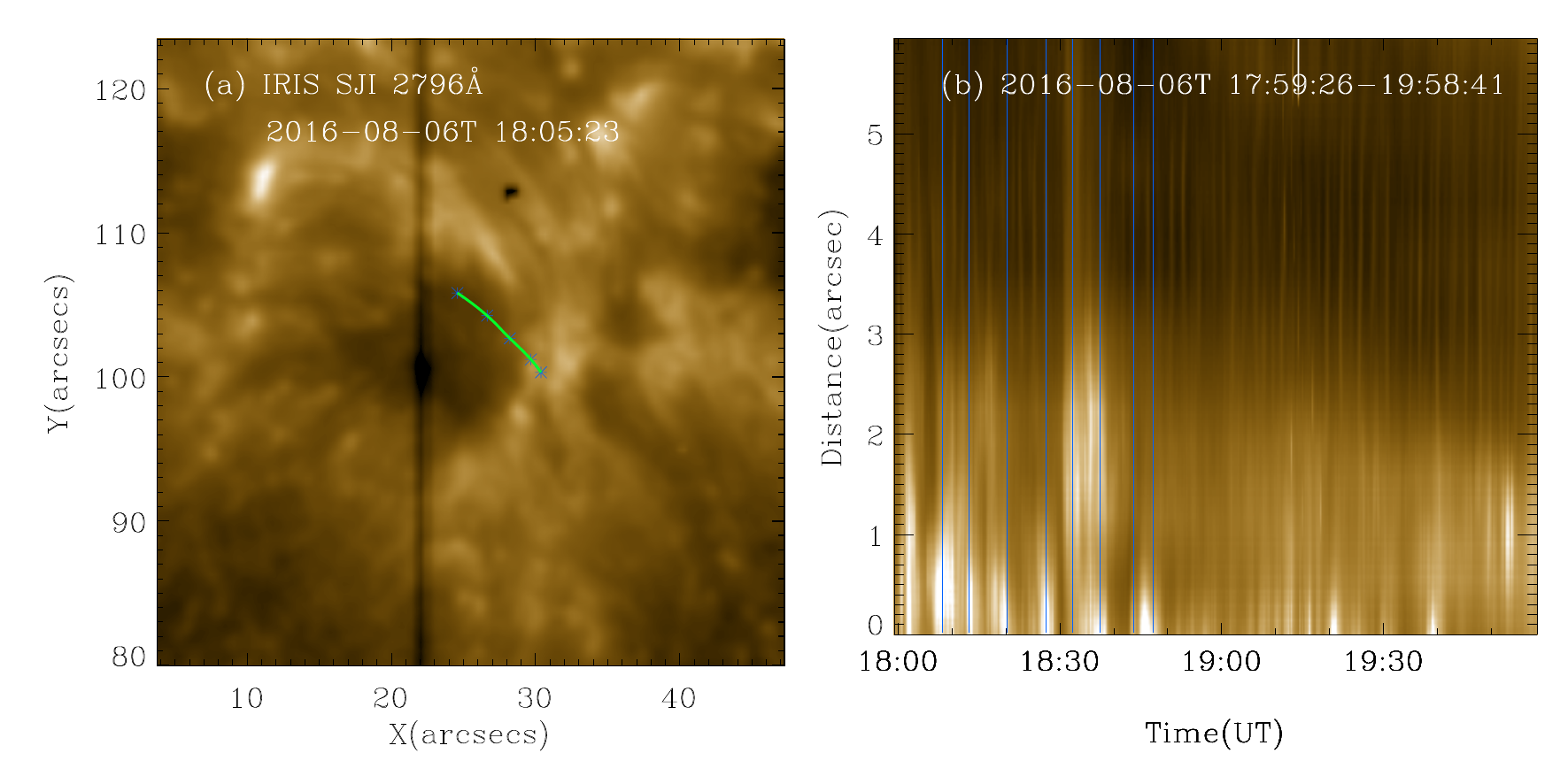}
\end{minipage}
\caption{ Oscillation behavior shown in the time-slice diagram of the evolution of the intensity at the jet base. Panel (a): IRIS SJI 2796 \AA\ image. The green curve indicates the location of the slit for the time-slice diagram. Panel (b): Time-slice diagram along the slit. The vertical blue lines were used to mark the moments of clear brightening.
}
\label{figure_5}
\end{figure*}

To further investigate the brightening phenomena, we chose the same slice position (S1 in Figure~\ref{figure_2}) in the SJI images and plotted the time-distance diagrams in 1400~\AA{} and 2796~\AA{} at this slice position (see Figure~\ref{figure_4} and Figure~\ref{figure_5}). The two time-distance diagrams show that the recurring brightening phenomena are presented throughout the two-hour observation period in both 1400~\AA{} and 2796~\AA{}, exhibiting a periodic behavior. We mark the peak time of the brightenings in the time-distance diagrams with vertical blue lines, and we found that the brightenings in 2796~\AA{} and 1400~\AA{} are consistently delayed for 2-3 minutes compared to the vertical green line in Figure~\ref{figure_2}(b). This indicates that the intermittent jets require some time to heat the upper solar atmosphere, and these brightening phenomena correspond to the response of intermittent jets in the lower transition region. \\

 No spectral data in this region are available for the IRIS observations, 

\subsection{Dopplergram of the intermittent jets}
 
In order to determine the line-of-sight (LOS) velocity of the intermittent jet material, we created Doppler diagrams of the plasma using the H$_{\alpha}$ 5 wavelength images from the GST observations. We calculated the center of weight of the H$_{\alpha}$ line profile at each pixel to estimate the Doppler shift relative to the reference line center. We averaged the entire observing FOV to obtain the reference line center (except in the region where the sunspot was located), and all the line profiles were corrected by comparing them with a standard H$_{\alpha}$ profile, obtained from the NSO/Kitt Peak FTS data \citep{su2016}.
 
\begin{figure*}[!htbp]
\begin{minipage}{\textwidth}
\centering
\includegraphics[width=170mm,angle=0,clip]{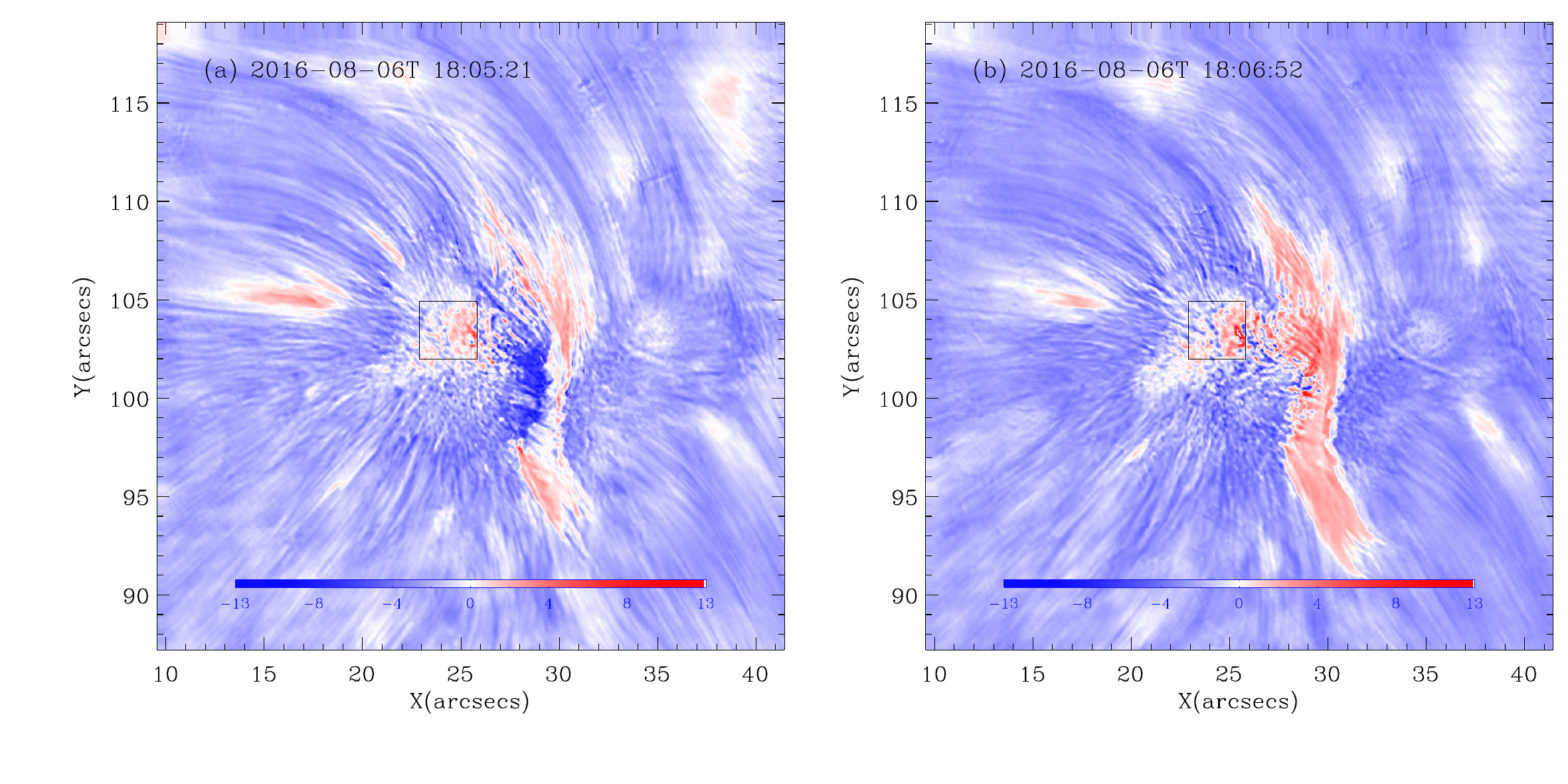}
\end{minipage}
\caption{LOS Doppler velocity maps obtained from GST/VIS. The black box shows the subtracted reference region. The two moments represent different patterns in eruption and fallback of intermittent jets.}
\label{figure_6}
\end{figure*}

Dopplershift maps presenting the LoS velocities are shown in Figure~\ref{figure_6} in blue and red for the blue and redshift motions. The jet exhibits a Doppler blueshift in Figure~\ref{figure_6} (a) and a Doppler redshift in Figure~\ref{figure_6} (b). This is consistent with the observations in H$_{\alpha}$ , in which the jet displays continuous upflow and downflow over 4 hours, indicating that the jet is intermittent. We found that the average upflow speed is up to $-13~\rm{km\,s}^{-1}$ and the average downflow speed is about $11~\rm{km\,s}^{-1}$ between 18:00 UT and 18:50 UT by using the shift of the central wavelength of the H$\alpha$ profile. 
 Contrast or cloud model methods for deriving the jet Dopplershifts would lead to upflows and downflows of -40 and 80$\;\rm{km\,s}^{-1}$, respectively.
 
\subsection{Spectral H$_{\alpha}$ line profile}

We normalized the intensity and exposure time of the H$_{\alpha}$ 5 wavelength images. It is worth noting that the reference H$_{\alpha}$ profile (averaged spectral profile of H$_{\alpha}$ in the quiet region) and the footpoint profiles are symmetric. 

\begin{figure*}[!htbp]
\begin{minipage}{\textwidth}
\centering
\includegraphics[width=150mm,angle=0,clip]{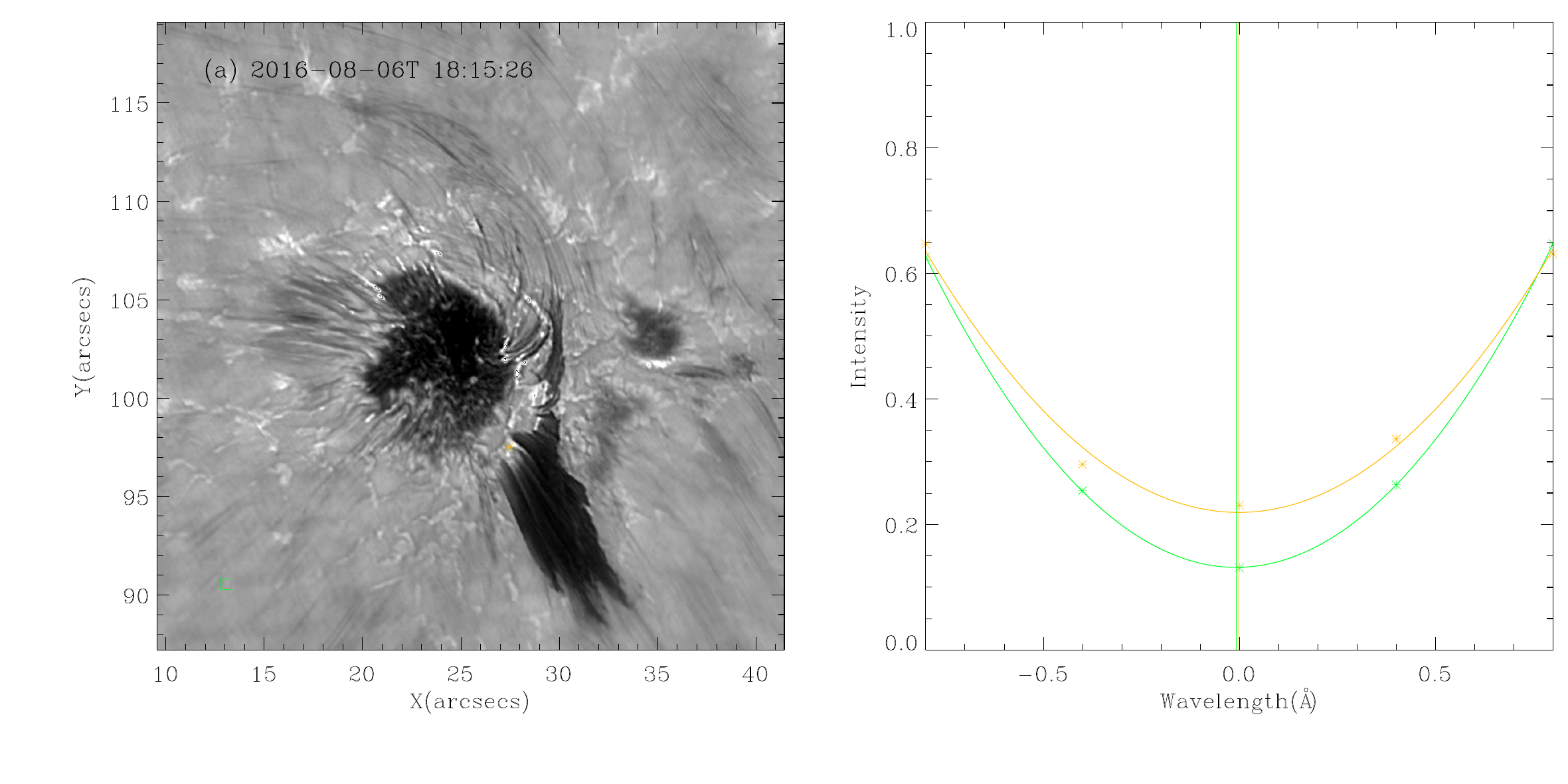}
\includegraphics[width=150mm,angle=0,clip]{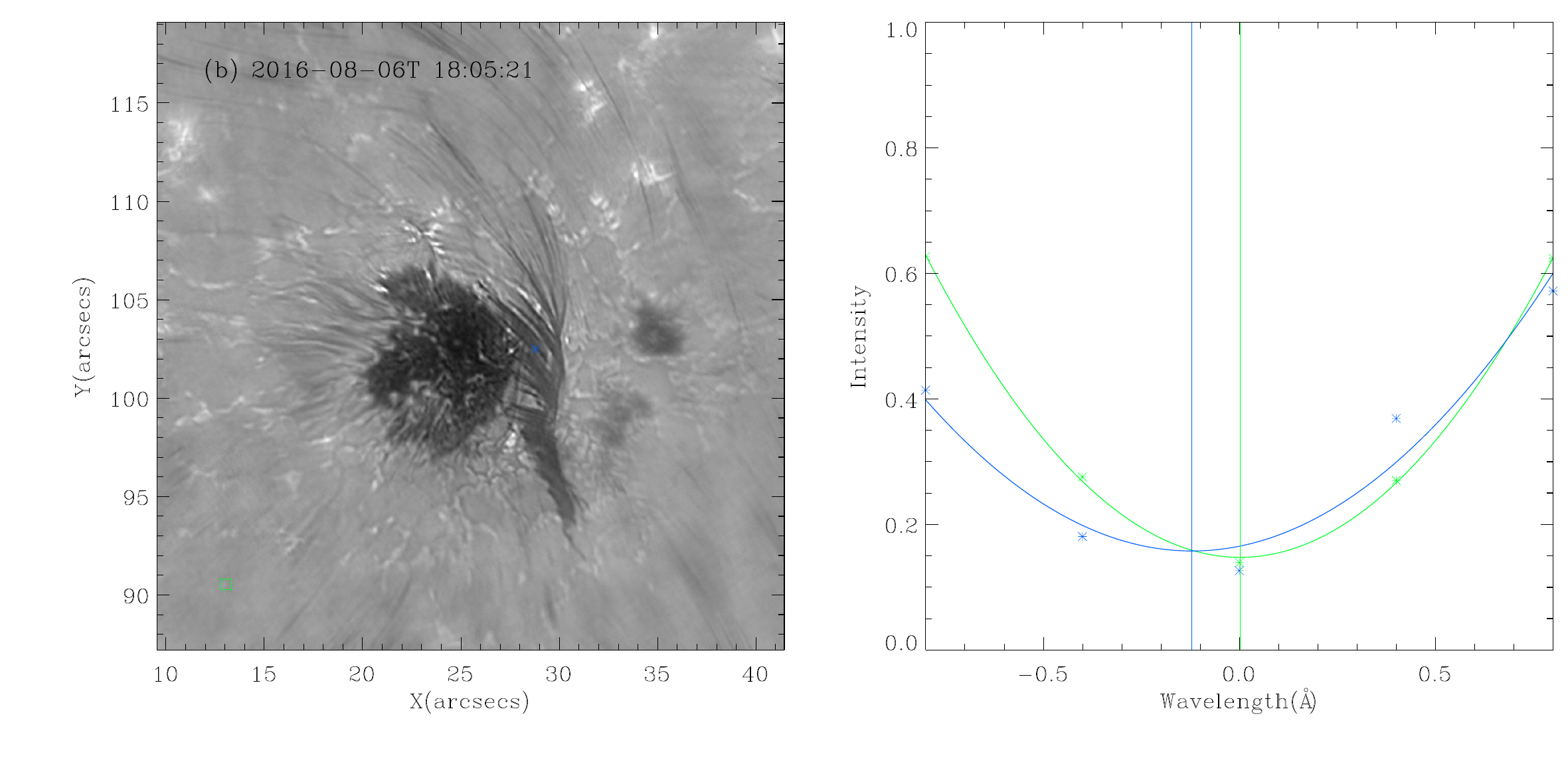}
\includegraphics[width=150mm,angle=0,clip]{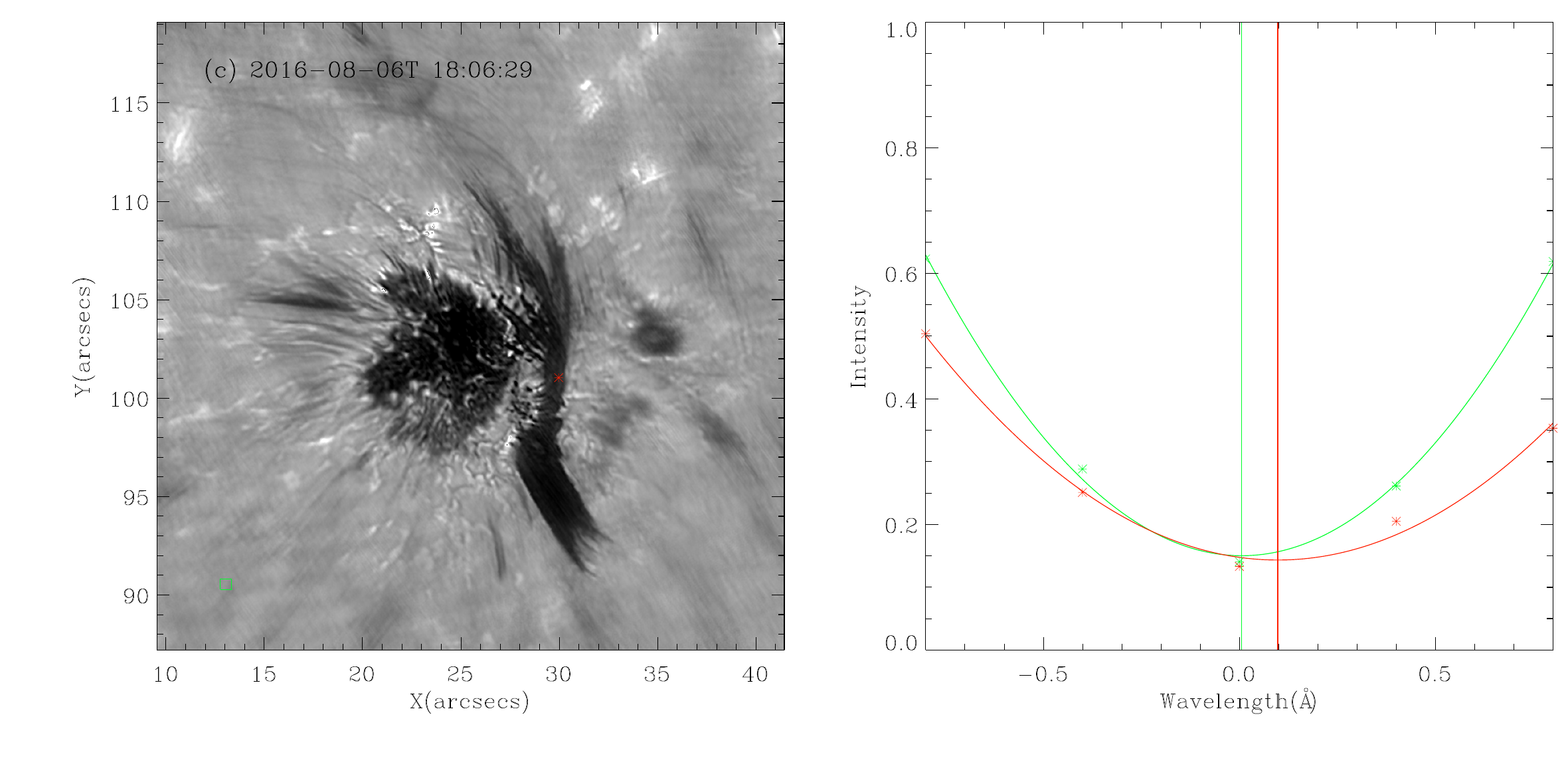}
\end{minipage}
\caption{Examples of normalized H$_{\alpha}$ spectral profiles in the jet and its footpoint (right panels). The top right panel corresponds to the profiles of the jet footpoint at 18:15:26 UT.   The normalized H$_{\alpha}$ spectral profile (yellow line) at the footpoint of the jet and reference profile (green line) are shown in the right panel. The footpoint (yellow dot) and the quiet region (green box) are marked in the image of the H$_{\alpha}$ +0.8 \AA{} in panel (a). In panels (b) and (c), the upflow (blue dot) and downflow (red dot) of intermittent jets are marked in the images of H$_{\alpha}$ +0.8 \AA{} 
at 18:05 UT and at 18:06 UT,
respectively. The normalized H$_{\alpha}$ spectral profile of the upflow (blue line), downflow (red line), and the reference profile (green line) are shown in the corresponding right panels. }
\label{figure_7}
\end{figure*}

The corrected spectral line profile of the quiet region was used as reference to analyze the footpoint. In Figure~\ref{figure_7} (a), we selected the footpoint $[x,y]=[617,355]$ (yellow dot) in the image of H$_{\alpha}$ +0.8~\AA{}. The green curve shows the spectral line profile of the quiet region, and the yellow curve shows the spectral line profile of the footpoint. Compared with the quiet region, the spectral line profile of the footpoint is elevated at the center of the H$_{\alpha}$ line, but there is no obvious change in the line wing, which may be because the footpoint is heated while the jet is mainly controlled by cold plasma, so that there is no obvious heating.

In Figure~\ref{figure_7} (b), we selected the upflow $[x,y]=[637,556]$ (blue dot) in the image of H$_{\alpha}$-0.8~\AA{}. The green curve shows the line profile of the quiet region, and the blue curve shows the line profile of the upflow. The spectral line profile of the jet in the process of eruption is blueshifted, that is, the direction of the eruption is toward the observer. In Figure~\ref{figure_7} (c), we selected the position of the downflow $[x,y]=[704,477]$ (red dot) in the image of H$_{\alpha}$ +0.8~\AA{}. The spectral line profile of the downflow is redshifted, that is, the falling direction is away from the observer.

\begin{figure*}[!htbp]
\begin{minipage}{\textwidth}
\centering
\includegraphics[width=170mm,angle=0,clip]{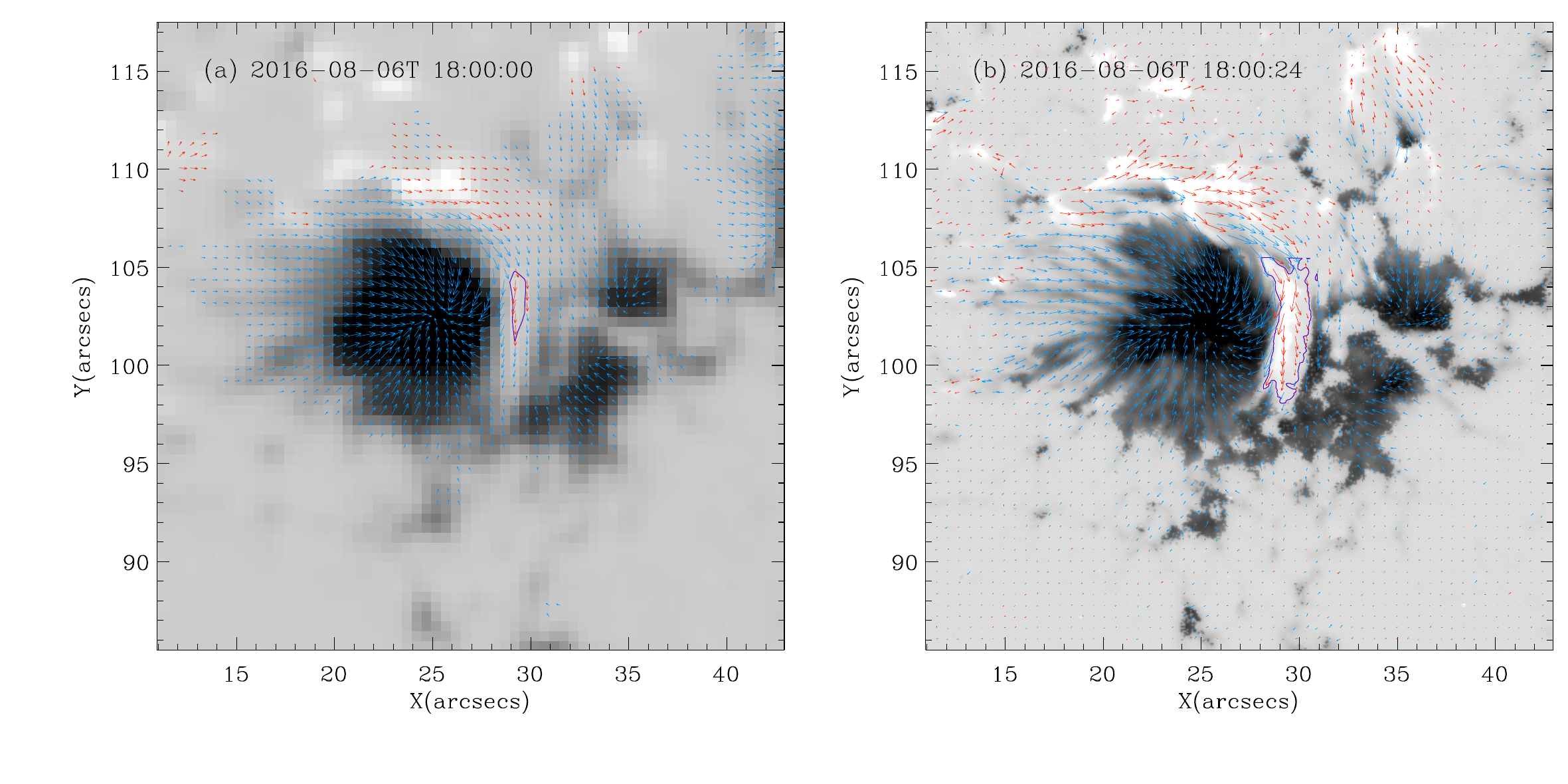}
\end{minipage}
\caption{The vector magnetic field from HMI and BBSO respectively. (a) Vector magnetogram of the FOV obtained from SDO/HMI. The vertical component of the vector magnetic field Bz in grayscale is overlaid with arrows. The red and blue arrows represent the strength and direction of positive and negative transverse fields, respectively. The purple contour indicates the boundary where the longitudinal field value is 0. The area is an important location for magnetic reconnection.
(b) Vector magnetogram of the FOV obtained from BBSO/NIRIS after removing the $180~\rm{^{\circ}}$  ambiguity in the transverse field. The vertical component of vector magnetic field, $B_z$, in grayscale is overlaid with arrows. The red and blue arrows represent the strength and direction of positive and negative transverse fields, respectively. The purple contour also indicates the boundary where the longitudinal field value is 0.}
\label{figure_8}
\end{figure*}

Using SDO/HMI Sharp cea vector magnetogram data, we overlaid the transverse field as arrows on the longitudinal field image to visualize the overall magnetic field information. In Figure~\ref{figure_8}(a), the negative-polarity magnetic field converges inward, while the positive-polarity magnetic field diverges outward, consistent with the direction of the positive and negative magnetic field lines. Additionally, in the regions in which positive and negative magnetic fields intersect, indicated by the near-parallel red and blue arrows, the magnetic field could imply 
a shearing behavior. This may provide the conditions for magnetic reconnection (see the area inserted with the purple contour in panel (a)).
Figure~\ref{figure_8}(b) shows the NIRIS vector magnetogram of the sunspot at the time of 18:00:24 UT obtained by applying the Milne-Eddington (ME) inversion to the Stokes profiles of the FeI 1565~nm doublet using the inversion code of J.\ Chae \citep{degl1992}. The azimuth component of the inverted vector magnetic field was processed to remove the $180 ^{\circ}$ ambiguity \citep{Leka2009}. A more detailed magnetic field structure appears that is consistent with the SDO/HMI vector magnetogram observations.

\begin{figure*}[!htbp]
\begin{minipage}{\textwidth}
\centering
\includegraphics[width=170mm,angle=0,clip]{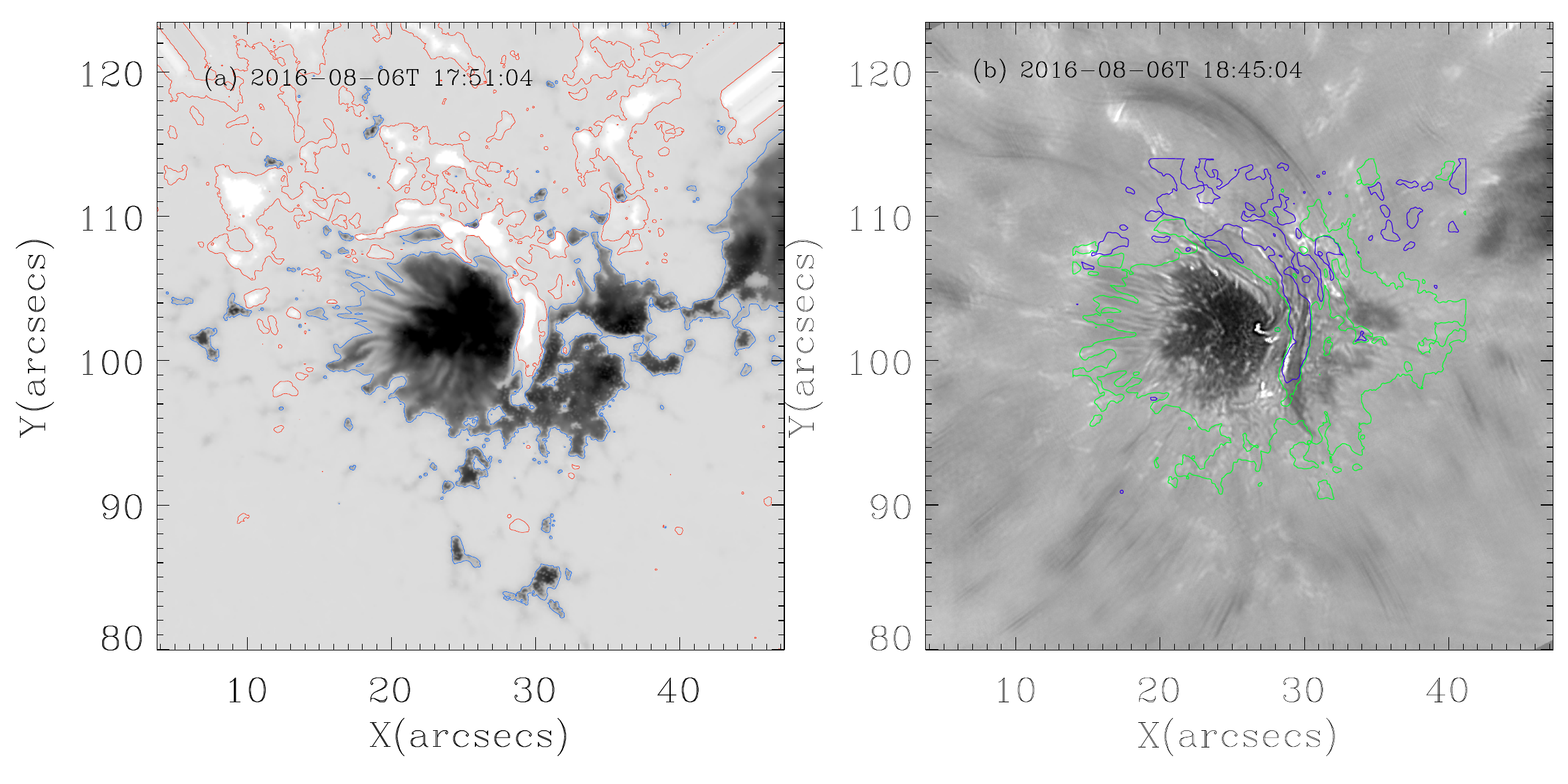}
\end{minipage}
\caption{(a) Longitudinal field obtained by BBSO inversion of the Stokes component. The red and blue line represent the magnetic field boundary contour of +20 G and -300~G, respectively.
(b) Observation image of the H$_{\alpha}$ blue-wing 0.8~\AA{} image superimposed on BBSO longitudinal field data. The purple and green lines represent the magnetic field boundary contour of +30 and -200~G, respectively. }
\label{figure_9}
\end{figure*}

Using the data from the BBSO four Stokes components, Figure~\ref{figure_9} (a) shows the image of the longitudinal field. The spot in the center of the field of view is a negative magnetic field, and there is a positive magnetic field in the middle of the spot. The jet occurs in the mixed polar magnetic field region. It may indicate the magnetic flux emergence, magnetic cancellation, and so on. In order to better determine the magnetic field environment of the jet, the high-resolution BBSO longitudinal magnetic field data were superimposed on the H$_{\alpha}$-0.8~\AA{} image. In Figure~\ref{figure_9} (b), the footpoint is the junction of positive and negative magnetic fields, near the magnetic neutral line. Compared with IRIS observations (see Figure~\ref{figure_3}), we found that the jet corresponds to the brightening phenomenon.

\begin{figure*}[!htbp]
\begin{minipage}{\textwidth}
\centering
\includegraphics[width=170mm,angle=0,clip]{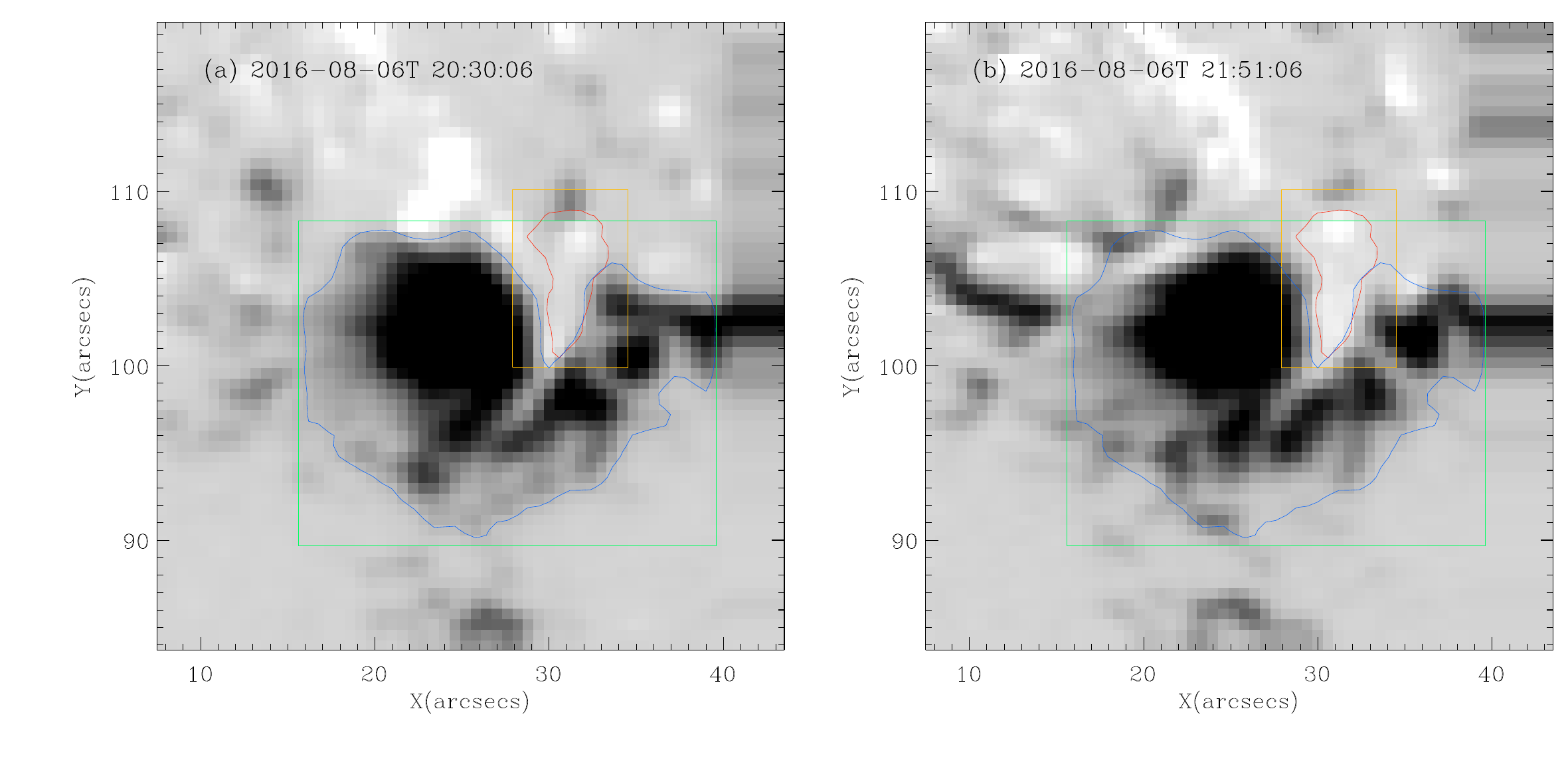}
\end{minipage}
\caption{Observation image of the HMI M45s longitudinal field. The region surrounded by red lines in panels (a) and (b) is the positive polar magnetic field evolution region 1 determined by the magnetic field value at 20:30:06 UT of +25 Gauss, and the blue line is the negative polar magnetic field evolution region 2 determined by the magnetic field value at 21:51:16 UT of -120 Gauss. The yellow and green box represent the rectangular region in which the positive and negative magnetic field evolution region, respectively, is located.}
\label{figure_10}
\end{figure*}

\begin{figure*}[!htbp]
\begin{minipage}{\textwidth}
\centering
\includegraphics[width=180mm,angle=0,clip]{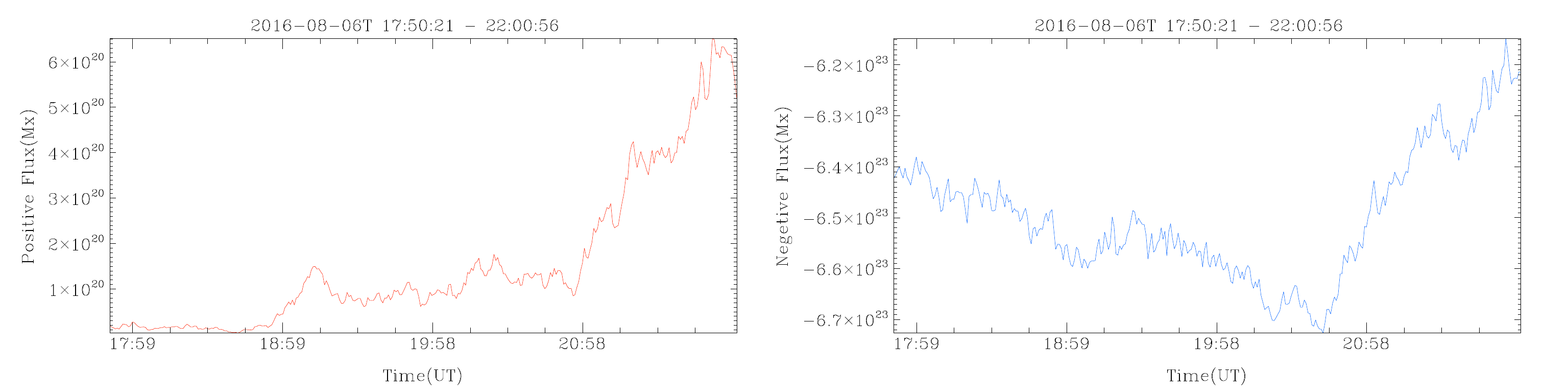}
\end{minipage}
\caption{Time-evolution diagram of positive and negative magnetic flux during the intermittent jet event. The red line represents the positive magnetic flux, and the blue line represents the negative magnetic flux.}
\label{figure_11}
\end{figure*}

In Figure~\ref{figure_10}, using the longitudinal field data of hmi.M\_45s, we selected two boundary values of positive and negative magnetic fields to determine the evolution region, and we then calculated the magnetic flux. In Figure~\ref{figure_11}, the positive magnetic flux continues to increase, which corresponds to the emergence of a positive magnetic field, while the negative magnetic flux first increases and then decreases, which may imply the cancellation of positive and negative magnetic fields. The emergence and cancellation of positive and negative magnetic fields may result in parallel and opposite-polarity magnetic field components, which may trigger magnetic field reconnection. This might trigger the intermittent jet.

\subsection{The movement of granulation}

In the evolution images of TiO band, we found that there is an extrusion movement of granulation right of the spot. The local correlation tracking (LCT) method was used to superimpose the horizontal flow velocity of the photosphere on the TiO image to show the motion of granulation. 

\begin{figure*}[!htbp]
\begin{minipage}{\textwidth}
\centering
\includegraphics[width=170mm,angle=0,clip]{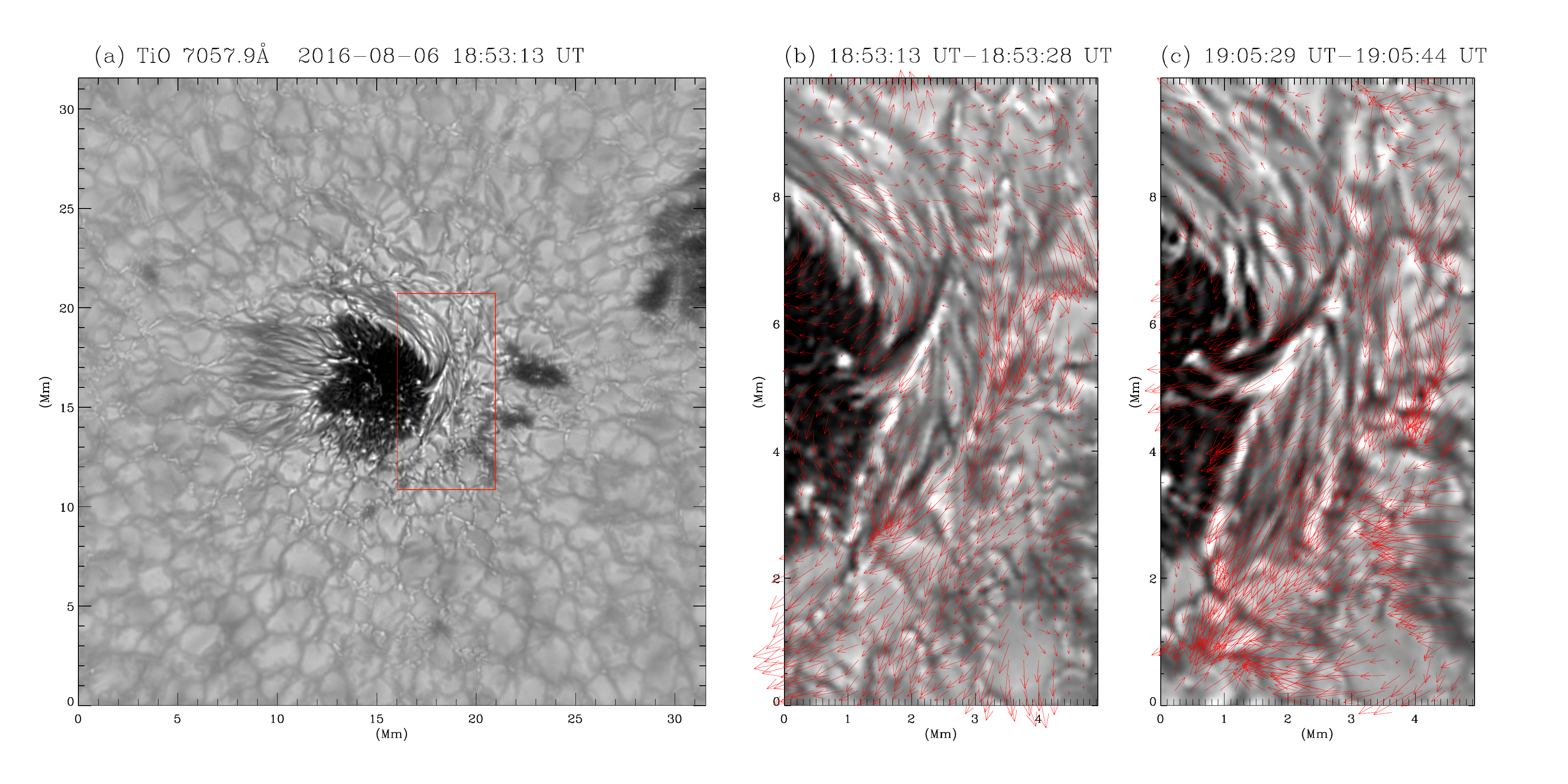}
\end{minipage}
\caption{Horizontal motion of the granulation in the photosphere. The red arrows in panels (b) and (c) represents the horizontal flow of the photosphere in the red box from panel (a). Between 18:53 UT and 19:05 UT, the whorl area around the west side of the sunspot extends toward the southwest, exhibiting elongated granules with a high horizontal velocity field.}
\label{figure_12}
\end{figure*}

Panels b and c in Figure~\ref{figure_12} represent the TiO emission in the area inserted in the small box drawn in panel a. This area is on the west side of the sunspot in its penumbra. In this area, the fibrils are not radial but turn counterclockwise around the spot. The fibrils and magnetic field lines are anchored in the penumbra. In panels b and c, elongated  granules lie in this area, and  around it lie fragmented granules with filigrees in the intergranules and black pores. The horizontal components of the velocity vectors are overlaid on the TiO maps. In a few points, the arrows converge (in panel b point (2,2.5), and (3.5,5). They move southwest to points (in panel (c)) (1,1) and (4,4). These locations move continuously away from the spot with time as  the area of th  elongated granules extends toward the quiet Sun.
This continuous motion might favor flux cancellation. The
 periodicity of the brightenings of around 5 minutes measured in this region (Figures 2, 4, and 5)  would be due to the p-mode oscillations of the photosphere that modulate   
 the radiation in this region.


\begin{figure*}[!htbp]
\begin{minipage}{\textwidth}
\centering
\includegraphics[width=170mm,angle=0,clip]{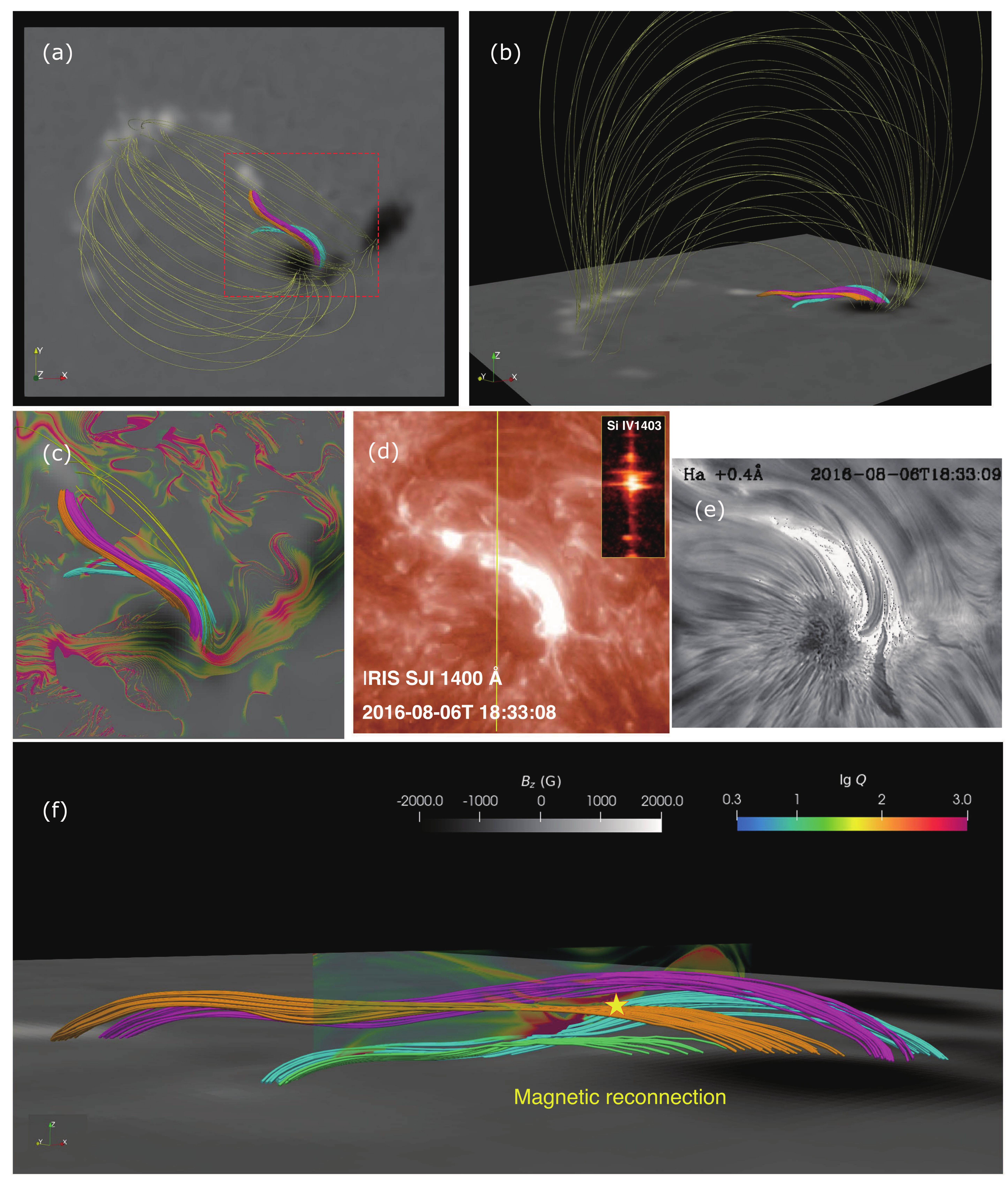}
\end{minipage}
\caption{3D magnetic configuration constructed with the NLFFF model. Panels (a) and (b) exhibit the top and side views of a few typical field lines, respectively. 
Panel (c) presents the corresponding zoom-in view and the bottom $Q$ distribution. The cyan and orange field lines might reconnect and lead to the longer pink field lines above and to the short green loops below. The long yellow lines are active region magnetic field lines.
Panels (d) and (e) exhibit the 1400~\AA\ and H$\alpha$-0.8 \AA\ images at 18:32~UT observed by the IRIS and GST, respectively. The yellow vertical line in panel (d) represents the IRIS slit, and the inset shows the spectra of Si IV 1403~\AA. Panel (f) shows the reconnection configuration and the distributions of the squashing degree $Q$.}
\label{figure_13}
\end{figure*}

\begin{figure*}[!htbp]
\begin{minipage}{\textwidth}
\centering
\includegraphics[width=170mm,angle=0,clip]{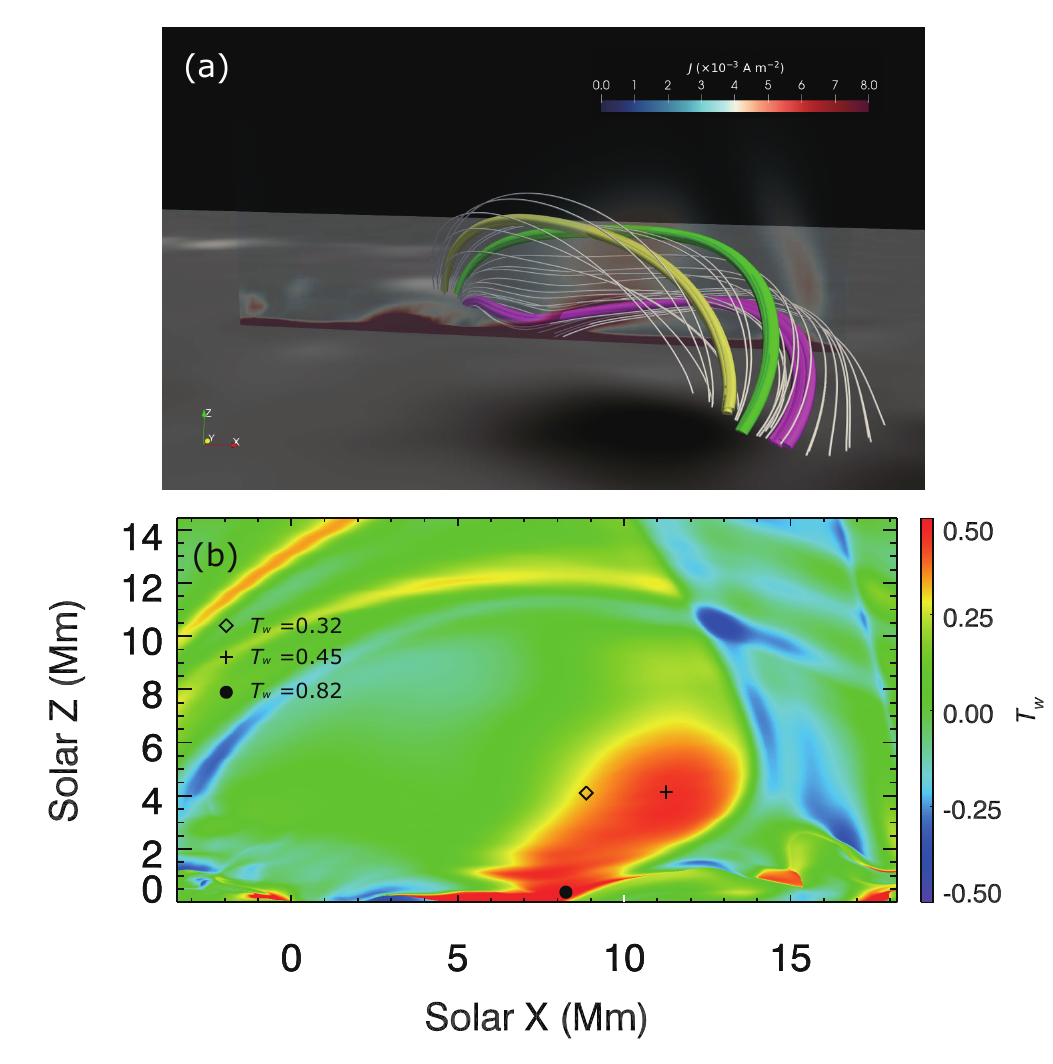}
\end{minipage}
\caption{Panels (a) and (b) display the typical magnetic field lines inside the electric-current channel and the twist map, respectively. The semitransparent vertical slice in panel (a) shows the distribution of the electric current. The purple, green, and yellow tubes in panel (a) are traced from the locations of the circle, plus, and rhombus in panel (b), respectively.}
\label{figure_14}
\end{figure*}

\section{3D magnetic field modeling}

To identify the magnetic structure of these intermittent jets, we reconstructed the 3D magnetic fields with the NLFFF extrapolation. The NLFFF extrapolation was carried out with the magnetofrictional (MF) method \citep{Guo2016t,Guo2016A} that is implemented in the framework of MPI-AMRVAC \citep{Xia2018, Keppens2020, Keppens2023}. The magnetofrictional model simplifies the MHD relaxation process by omitting the effects of inertia, gravity, and pressure gradients, and the velocity is assumed to be proportional to the local Lorentz force. Consequently, the magnetic field evolution is governed by the magnetic induction equation, and the relaxed state approaches a force-free state \citep{Yang1986}. More details regarding the implementation and numerical schemes of the magnetofrictional model in the AMRVAC framework can be found in \citet{Guo2016A, Guo2016t}.

We adopted the vector magnetogram (hmi.B\_720s) at 18:24~UT observed by the SDO/HMI to reconstruct the coronal magnetic field. To ensure that the observed vector magnetic fields in the photosphere adhere to the assumptions of the boundary condition of the NLFFF model, we performed some preprocessing steps, including correcting projection effects \citep{Guo2017} and removing the Lorentz force and torque \citep{Wiegelmann2006}. The initial magnetic field for the magnetofrictional model was derived using the Green function method \citep{Chiu1977} with the $B_{z}$ component. The processed vector magnetic fields ($B_{x}$, $B_{y}$, and $B_{z}$) were imposed on the inner ghost layer of the bottom boundary, and the values in the outer ghost layer were provided by zero-gradient extrapolation. After 60000 iterations of the magnetofrictional relaxation, the force-free metric (for more details, see \citet{Guo2016A}) decreased to half the value of the initial potential field, while the electric current doubled. Figure~\ref{figure_13} illustrates the results after 60000 iterations of the magnetofrictional relaxation.

Figures~\ref{figure_13} (a) and \ref{figure_13} (b) display the top and side views of the reconstructed 3D magnetic fields, respectively. The yellow tubes represent the overlying background fields, and the remaining tubes that are highly sheared (cyan, pink, and orange) delineate the topological structure related to the areas in which intermittent jets take place. 
A good consistency can be seen between the reconstructed magnetic field lines and the observations, as highlighted in the middle panels of \ref{figure_13}. On the one hand, the pink tubes exhibit a morphology resembling the bright structures observed by the IRIS/SJI 1400~\AA. On the other hand, the cyan tubes are similar to the observed fibrils in H$_{\alpha}$ red-wing images. These results suggest that our NLFFF model almost retrieves the magnetic-field structures of the observations to a great extent.

Hereafter, to understand the underlying mechanisms resulting in the observed jets, we calculate the distribution of the squashing degree $Q$ \citep{Priest1995, Demoulin1996} with the method proposed by \citet{Scott2017}. The regions of $Q\gg 2$ depict the areas in which magnetic connectivity undergoes drastic changes, commonly refereed to as quasi-separatrix layers (QSLs), which outline the favorite places for magnetic reconnection \citep{Titov1999, Aulanier2010, Janvier2013, Guo2013, Li2022, Zhong2021, Guo2023}. Figure~\ref{figure_13} (f) shows  selected magnetic fields around the QSLs. It shows that the cyan and orange field lines form an X-shaped configuration featured by the areas with high $Q$ values. The reconnection between them could produce longer pink S-shaped field lines above and the green shorter loops below. 
A strong squashing factor  (Q) value is detected at the base of the jet.  This will allow for the field lines to move their direction from north to south and explains the southern component of the jets. For instance, the yellow lines in panel (c)  with southern ends in the high-value Q factor can reconnect toward the south instead of joining the positive northern polarities.

Figure~\ref{figure_14} (a) illustrates the configuration of the core magnetic field of the jets. The field lines traced from the electric current channel are sheared and twisted. To quantify the degree of twist, we computed the distribution of the twist number in the same plane as the electric current intensity, denoted as $T_{w}$, using the open-source code implemented by \citet{Liu2016}. As depicted in Figure~\ref{figure_14}b, the high-$T_{w}$ region forms a quasi-circular shape, corresponding to the electric current channel in Figure~\ref{figure_14}a. Furthermore, the twist number of the purple field line in Figure~\ref{figure_14}a can reach a value of 0.82, which is formed due to magnetic reconnection illustrated in Figure~\ref{figure_13}. Taken as a whole, the investigation of the 3D magnetic fields indicates that magnetic reconnection might cause the observed jets. This agrees with our observations.

We checked the IRIS Si IV spectra along the slit indicated in \ref{figure_13} (d). Two mini flares were registered at this location. 
During one flare at 18:33 UT, bilateral flows occur (\ref{figure_13} (d) inset). This indicates reconnection between the strands of the jet \citep{Ruan2019}.


\section{Discussion and conclusions}
We reported coordinated observations with the GST at BBSO , SDO/HMI, and IRIS SJI of active region NOAA 12571 on August 6, 2016. The observed event shows intermittent jets that occur right of a negative-polarity sunspot.

 We obtained the following results:
\begin{enumerate}
 \item In the H$_{\alpha}$ +/-0.8\AA{} , we found a persistent jet right of the sunspot that lasted for up to 4 hours. The intermittent jets exhibited outward-diverging dark absorption features. The time-distance diagram shows that the peak of the jet has clear periodic eruption characteristics (5 minutes) during 18:00 UT-18:50 UT. 
 
 \item We also observed a periodic brightening in the transition region during the jets, which was reflected in the time-distance diagram. This may be a response of the intermittent jets in the higher solar atmosphere.
 
 \item By calculating the Doppler velocities of the jets, we found alternating redshifts and blueshifts during the eruption that correspond to the intermittent eruptive nature of the jets. The average  velocities of the upflow and downflow are $-13.47~\rm{km\,s}^{-1}$ and $11.11~\rm{km\,s}^{-1}$, respectively, when we consider the displacement of the central wavelength in the H$\alpha $ profiles.

 \item  Compared to the quiet region of the Sun, we found that the spectral line profiles at the footpoint showed an 
  intensity increase at the line center, but no significant changes in the line wings. This indicates   prolonged heating at the footpoints.

\item In the vector magnetograms of the intermittent jets, we observed a strong shear behavior in the magnetic field near the neutral line. This provides a favorable magnetic environment for magnetic reconnection.

\item By overlaying the BBSO longitudinal magnetic field onto the H$_{\alpha}$ and the IRIS SJI images in three wavelengths, we determined that the footpoints were located on the neutral line, corresponding to the brightening in the transition region and lower chromosphere. The magnetic flux evolution diagram shows that positive magnetic flux emerges continuously, and  negative magnetic flux first increases and then decreases, indicating that the magnetic field has a long-duration emergence and cancellation behavior.

\item The evolution image in the TiO wavelength shows horizontal motions of the granulation at the location of jets. Granules intrude in this place. This may contribute to the compression of opposite-polarity magnetic fields and might trigger magnetic reconnection.

\item The magneto-topology analysis for the 3D NLFFF model confirms the possibility that magnetic reconnection in the corona is the main mechanism for the production of intermittent jets. This is also confirmed by spectroscopy data with bilateral flows.

\item The magnetic field lines containing the jets are anchored in the mixed magnetic polarity channel inside the negative-polarity spot.  

\end{enumerate}
  
  
Magnetic reconnection has been widely accepted as the driving mechanism of most solar eruptive events. Many small-scale events in the solar atmosphere, such as solar jets \citep{Asai2001,Mulay2016,Raouafi2016,Shen2021}.
Many observational events can be explained by magnetic reconnection between preexisting open magnetic field lines and newly emerging magnetic fields, and the magnetic cancellation driven by newly emerging magnetic bipoles can be regarded as slow magnetic reconnection in the lower solar atmosphere \citep{Wang1993,Jiang2000}. The appearance of microflares, the converging form of EUV jets, and the relation between their footpoints and annular flare brightenings all serve as evidence of magnetic reconnection in jets \citep{Shibata1998}. 

In this paper, the intermittent jet occurred in a mixed-polarity region at the boundary 
negative magnetic fields. By overlaying the magnetic field onto the observed images of the jets, we found that jets were located near the neutral line. The evolution of the magnetic flux showed a prolonged magnetic flux emergence and magnetic cancellation. This provides a favorable magnetic environment for the intermittent jets that continuously drive the ejection of plasma material. 
Furthermore, in the observations of the photosphere, we found horizontal motion of the granulation in the eruption area.



The relative motion between granulation and  the magnetic field along the neutral line may lead to flux cancellation, which is favorable for initiating jets \citep{Tian2017,Yang2015,Li2020}. The magnetic field lines could be submitted to the oscillations present in the sunspot and its environment. 
We showed some evidence that magnetic reconnection could be the mechanism that triggers the  intermittent jets,
although to confirm the temporal nature of our proposed mechanism, a
time series of extrapolations is required.\\
 

Previous studies have investigated the triggering of repetitive jets by magnetic emergence and cancellation \citep{Zhang2014,Chae1999,Liu2016,Zeng2013}, but most events lack a periodicity. However, \citet{Ning2004,Doyle2006,Chandra2015} discovered repetitive eruptive events in the transition region with quasi-periodic characteristics. \citet{Wang2021,Wang2023} analyzed jets that also exhibited quasi-periodic features of approximately 5 minutes, and \citet{Hong2022} observed long-duration microjets that displayed a quasi-periodic behavior of approximately 5 minutes. 
Additionally, \citet{Chen2006} used MHD simulations to demonstrate the scenario of periodic magnetic reconnection modulated by p-mode oscillations in eruptive events.


\begin{acknowledgements}
Guiping Ruan and Qiuzhuo Cai acknowledge the support by the NNSFC grant 12173022, 11790303 and 11973031. We gratefully acknowledge the use of data from the Goode Solar Telescope (GST) of the Big Bear Solar Observatory (BBSO). BBSO operation is supported by US NSF AGS-2309939 and AGS-1821294 grants and New Jersey Institute of Technology. GST operation is partly supported by the Korea Astronomy and Space Science Institute and the Seoul National University.
We thank Dr.\ S.\ Poedts for reading and bringing some fruitful comments to the manuscript. We are grateful to Shuhong Yang for the discussion on the manuscript. W. Cao acknowledges support from US NSF AST-2108235 and AGS-2309939 grants. B.S.\ and J.H.G.\ acknowledge support from the European Union’s Horizon 2020 research and innovation programme under grant agreement No 870405 (EUHFORIA 2.0) and the ESA project "Heliospheric modelling techniques“ (Contract No. 4000133080/20/NL/CRS). We thank SDO/HMI, SDO/AIA teams for the free access to the data.  
 
\end{acknowledgements}


\bibliography{reference.bib}
\bibliographystyle{aa}

\end{document}